\documentclass[a4paper,11pt,final]{article}

\usepackage{graphicx,amsmath,amssymb} 
\usepackage{epsfig}

\begin{document}

\title{\ \\ \LARGE\bf  Scale--invariance of ruggedness measures in fractal fitness landscapes}

\author{Hendrik Richter \\
HTWK Leipzig University of Applied Sciences \\ Faculty of
Electrical Engineering and Information Technology\\
        Postfach 301166, D--04251 Leipzig, Germany. \\ Email: 
hendrik.richter@htwk-leipzig.de. }

\maketitle

\begin{abstract}
The paper deals with using chaos to direct trajectories to targets and analyzes ruggedness and fractality of the resulting fitness landscapes.  The targeting problem is formulated as a dynamic fitness landscape and four different chaotic maps generating such a landscape are studied. By using a computational approach, we analyze properties of the landscapes and quantify their fractal and rugged characteristics. In particular, it is shown that ruggedness measures such as correlation length and information content are scale--invariant and self--similar.

\end{abstract}

\section{Introduction}
A considerable number of scientific  application domains promote a mathematical modelling which is centered on  a quality information that projects over decision variables. For some application domains the decision variables may be payoff attached to evolutionary games, as in evolutionary game theory defining payoff landscapes~\cite{brede11}, or trading strategies as in financial market analysis, which entails  profit landscapes~\cite{groe12}. Further examples of  decision variables are
conformations of  molecular entities or spatial positions of interacting molecules
as in the theory of spin glasses or folding and energy relaxation in proteins and nucleic acids, which uses energy landscapes~\cite{ban05,on97,wal04}, or cluster structures in large data analysis, which employs cost landscapes~\cite{gia02}, or control variables of electromagnetic fields as in quantum dynamics, which defines
control landscapes~\cite{palao13,rabitz14}.
Lastly, and arguable most prominently, the variables may be
 genotypes or search space elements, as in evolutionary biology and evolutionary computation defining fitness landscapes~\cite{gav04,stad96,richengel14}.

The common denominator of all these modelling paradigms is that they permit the geographical metaphor of a landscape as the quality information can be interpreted as height  with high (or low) values  forming  peaks (or valleys). There is a further common feature in all these landscape approaches. Deriving useful information from the landscape requires a notion of how many maxima (or minima) there are, and how they are distributed and accessible. This is connected with studying if there are regularities, pattern and invariances over varying  dimensions and/or scales. The second common feature frequently culminates in defining metrics of the landscapes, for instance by landscape measures or other invariant quantities such as fractal dimensions. These landscape metrics have the advantage to be applicable to landscapes of higher dimensions for which the geographical metaphor might be rather hard to interpret or even outright meaningless. 

The application domain considered in this paper is analyzing and controlling chaotic systems. More specifically, control is studied  that uses chaos to direct trajectories to targets, and a landscape modelling is employed. Targeting control exploits specific properties of chaos, namely orbit density and ergodicity, and employs  tiny perturbations to system parameters for making the trajectory visiting pre\-selected points on the chaotic attractor. Consequently, the perturbations to the system parameters are the decision variables, while the success of the control is measured as the distance between intended target  and  actual hit.  Thus, finding the targeting control can be interpreted as solving an optimization problem and it has been shown that this optimization problem poses a dynamic fitness landscape~\cite{rich12}. In this paper these
 results are extended from dynamic 1D landscapes to dynamic 2D landscapes. Another addition is that the focus is now put on analyzing the fitness landscapes and particularly on measuring if and to what extend the landscapes are rugged and fractal. There are several works dealing with fractal landscapes~\cite{hosh98,mac06,sork91,wein93,zelink14} or rugged landscapes~\cite{kauf87,stad96,rich08,richengel14}. This also applies to the field of analyzing and controlling chaotic systems. For instance, there are escape time landscapes~\cite{boll05}, which have shown to possess fractal properties~\cite{leit13}. These landscapes originate from considering which initial states escape the chaotic transients and how long it takes. In other words, the decision variables are  initial states, fitness is transient time. Although most of these works acknowledge that rugged and fractal are linked, and fractality   is even described as an ``extreme case''~\cite{leit13} of ruggedness, 
the interplay between these two landscape characteristics  is still somewhat obscure. Do landscape measures evaluating ruggedness show self--similarity and other fractal properties by varying the scales upon which they are calculated? Does ruggedness come to an end if the scale of the configuration space variables goes to zero? Such questions are raised  and by using the example of targeting control, 
general relationships between ruggedness and fractality of landscapes are studied.

The outline of the paper is as follows.  In Sec.  \ref{sec:chao}, properties of chaos and their significance for targeting control are briefly recalled, and the dynamic landscape models are introduced.  Ruggedness and fractality of landscapes are discussed in Sec. \ref{sec:land}, and methods for numerically evaluating these landscape characteristics are reviewed. Also,  a definition of fractality of fitness landscapes is proposed.
Sec.  \ref{sec:exp} reports numerical experiments using four different dynamical systems showing chaos. For these systems, the targeting landscapes are analyzed, and their fractal dimensions are calculated. It is further demonstrated that ruggedness measures such as correlation length and information content exhibit scale--invariance and self--similarity. 
The paper closes with Sec. \ref{sec:con} giving a summary and concluding remarks. 

\section{Chaotic behavior and targeting} \label{sec:chao}

Consider a discrete--time dynamical system  \begin{equation} x(k+1)=f(x(k),p) \label{eq:basic} \end{equation} with the state variable $x \in \mathbb{R}^n$,  the discrete time variable $k$ of  an integer  time set $\mathbb{N}_0$,  and a map $f$  describing how the next state $x(k+1)$ is generated from the current state $x(k)$. The  system (\ref{eq:basic}) may additionally depend on a parameter vector $p \in \mathbb{R}^m$.  
Assume that the values of the parameter vector $p$ can be adjusted with time and hence be interpreted as a (time--dependent) control input $p=p(k)$. 
Further suppose that for some nominal values $p=\bar p$ and initial states $x(0)$, the trajectories of the system   (\ref{eq:basic}) are chaotic. Deterministic 
chaos implies several properties.
\begin{enumerate}
\item The system trajectory is highly sensitive against tiny perturbations of the initial state $x(0)$ and/or the control input $p(k)$. By applying such tiny perturbations, exponential divergence of nearby trajectories is caused. 

\item Chaotic trajectories are locally  not stable (in the sense of Lyapunov), but globally form a bounded and closed subset of the state space $\mathbb{R}^n$. No trajectory starting from this subset can escape it. The bounded and closed  set built by the chaotic trajectory is called the chaotic attractor $A_R$, and has but for exceptional cases a fractal dimension. 

\item Chaos is connected with orbit density implying that the trajectory comes arbitrarily close to almost all points embedded in the chaotic attractor. From the perspective of dynamics, dense orbits mean that the chaotic attractor is ergodic in the sense that almost all points in any subset of the attractor eventually get revisited  after a certain and finite time interval. 
\end{enumerate}
These properties of chaotic systems make it possible to direct trajectories to targets on the attractor $A_R$ by   
  bounded control perturbations 
\begin{equation} \label{eq:limit} \mathcal{P}=\{p\in \mathbb{R}^m | \|p-\bar{p} \|_\infty \leq \eta \} \end{equation}
with $\eta>0$ being a small constant.  Such a control policy is called targeting and directs the trajectory from any initial state $x(0)$ on a chaotic attractor $A_R$ within $\tau$ time steps to the neighborhood of any target point $\bar{x}$ on the same attractor~\cite{ip02,pas95}. Thus, we intend to achieve   $\| x(\tau)-\bar{x} \| \leq \alpha$
with  $\alpha \geq 0$ being another small constant. 

Whereas in principle the control input $p$ could have different values for each time step $k$,  targeting of chaotic systems is achieved by  using the control perturbation only once at the initial time $k=0$:
\begin{equation} \label{eq:steurfolg2}p(0)=p, \quad p(k)=\bar{p}, \quad k\geq 1. \end{equation}
Observing (\ref{eq:steurfolg2}) and denoting the multiple application of (\ref{eq:basic}) by $f \left(f(x(0),p),\bar{p}\right)=f^2(x(0),p)$ and so on, we can write  the fitness function of the targeting problem as
\begin{equation} \label{eq:opttarget}j(p)=  \left\|f^\tau(x(0),p)-\bar{x} \right\|. \end{equation} It accounts for the distance between the target $\bar{x}$ and state $x(\tau)$ after $\tau$ applications of the map $f$  and employing the perturbation $p$.
We interpret this function as a fitness landscape over the configuration space $\mathcal{P}$. According to the definition  (\ref{eq:limit}), the neighborhood structure of the landscape is inherent by the metric of real--valued vector spaces.   

For a given and fixed target time $\tau$, the fitness landscape (\ref{eq:opttarget}) is static. However, in view of the orbit density and ergodicity, it might be interesting to ask how the problem changes if the target time were to vary. Because of the ergodicity of the chaotic attractor, it might be favorable to wait (or to speed up) for one or more time steps to closer hit the target. By using the time variable $\kappa \in \mathbb{N}_0$ for varying $\tau$, we get a dynamic fitness landscape \begin{equation} \label{eq:opttargetdyn}j(p,\kappa)=  \left\|f^\kappa(x(0),p)-\bar{x} \right\|. \end{equation} 
Eq. (\ref{eq:opttargetdyn}) is a dynamic distance function accounting for the distance between the target $\bar{x}$ and state $x(\kappa)$ and allows us to  analyze the effect of the control perturbations $p$ on variable target time $\kappa$. 
    This dynamic fitness landscape is studied with respect to ruggedness and fractality in the next section. 

\section{Landscapes, ruggedness, and fractality} \label{sec:land}
A static fitness landscape is defined by the triple $\Lambda_s=(\mathbb{X},n,f)$~\cite{stad96,richengel14}. In this definition $\mathbb{X}$ is a configuration space made up by a finite or infinite number of configurations, $n(x)$ is a neighborhood structure sorting $\mathbb{X}$ by defining what is next to each $x \in \mathbb{X}$, and $f(x)$ is a fitness function giving every $x \in \mathbb{X}$ a proprietary quality information. The configuration space can be interpreted as being formed of the decision variables. For the static fitness function of the targeting problem (\ref{eq:opttarget}), we hence obtain the landscape $\Lambda_s=(\mathcal{P},d,j)$, where the space of control perturbations $\mathcal{P}$ is the configuration space with the Euclidean distance function $d$ serving as  neighborhood structure, and the static distance function $j(p)$ of the targeting problem (\ref{eq:opttarget}) is fitness. A dynamic fitness landscape can be defined by the quintuple $\Lambda_D=(\mathbb{X},n,\mathcal{K},F,\phi)$,~\cite{rich14a}, where $\mathbb{X}$ and $n$ are the same as in the static case, $\mathcal{K}$ is a time set of an integer time variable, $F$ is the set of all fitness functions $f(x,\kappa)$ in time $\kappa \in \mathcal{K}$, and the transition map $\phi$ describes how the fitness function changes over time.
For the dynamic fitness function of the targeting problem (\ref{eq:opttargetdyn}), we consequently obtain the landscape $\Lambda_d=(\mathcal{P},d,\mathbb{N}_0,\mathcal{J},\phi_j)$, where $\mathcal{J}$ comprises of all possible $j(p,\kappa)$ and $\phi_j$ can be constructed iteratively via the dynamical system (\ref{eq:basic}).

Although ruggedness of landscapes is an important topic, there are some difficulties in how to define it.
Clearly, the intuitive meaning  is that a rugged landscape possesses (apart from the global maximum) a substantial number of local maxima. 
This goes along with the landscape metaphor which depicts a structure equipped
``with many peaks, ridges and valleys'', as Kauffman and Levin~\cite{kauf87} put it introducing the concept of ruggedness. Unfortunately, such an intuitive meaning is hardly sufficient to establish free from much debate whether or not a given landscape is rugged. It has been suggested by Palmer~\cite{palmer91}  to define ruggedness by the property of the number of local maxima scaling at--least--exponential with the dimension of the configuration space. However, such a definition is prone to the limitation that focusing on the maxima alone does  possibly not fully catch the intuitive meaning of ruggedness as other landscape features are also influential, for instance the distribution of the maxima and their accessibility as expressed by the correlation between height and width of the maxima or the abundance of evolutionary paths leading to the maxima.   Also, using the definition to evaluate ruggedness requires either a designed landscape with an {\it a priori} known number of maxima, or to numerically count this number. Both  may be  awkward. Moreover, it is only applicable if there is a family of landscapes with varying dimension for which the at--least--exponential growth in the number of maxima might be established.  

 A kind of remedy came with landscape measures for calculating a numerical quantity of ruggedness~\cite{wein90,mun15,vassi00,stad96}.  Amongst the different groups of landscape measures, schemes based on random walks on the static landscape became particularly useful. These schemes work by recording the fitness value for each step of a random walk of length $T$  and forming  the sequence \begin{equation}  \mathcal{S}=(j(p_0),j(p_1),\ldots,j(p_{T-1}) ) \label{eq:seq} \end{equation} from walking $j(p)$. Thus, a series of neighboring fitness relations is obtained. Assuming that the landscape is isotropic, these neighboring fitness relations account for general changes in fitness across the landscape. For a dynamic landscape $j(p,\kappa)$, a sequence $\mathcal{S}(\kappa)$ can be obtained for each dynamic instance. Post--processing the sequence (\ref{eq:seq}) gives rise to numerical quantities whose values can be interpreted as measures of ruggedness. However, whether or not exceeding or falling below a certain limit value indicates a rugged landscape depends on several factors including the conditions of executing the random walk. As 
Stadler~\cite{stad96} rightly observed, ``it  seems  to be rather contrived to invoke a stochastic process $[$i.e a random walk$]$  in order to characterize a given function.'' Hence, ruggedness measures based on random walks are clearly suitable for comparing different settings of the same or similar landscapes. For generally defining ruggedness of a  landscapes, such measures are not very useful.

Thus, and despite the limitations discussed above, we may see the number of local maxima  as  
a defining property of ruggedness. Hence, for evaluating landscapes, it may be useful to have a lower and upper limit of ruggedness. The lower limit of ruggedness is easy to state. There  are  zero local maxima, but one global maximum, also known as Mt. Fuji landscape~\cite{richengel14}. The question of the upper limit of ruggedness can be addressed by the fractality of landscapes, employing an  ``intimate connection  between roughness and fractality'', as Alves and Hansmann~\cite{alv00}  phrased it  (albeit substituting ruggedness with roughness, synonymously, we may presume).  Consider a landscape with given configuration space, dimension and ruggedness. Now let the number of local maxima go up to increase ruggedness. For a continuous (real--valued) configuration space with an infinite number of possible configurations, the number of local maxima can be arbitrarily high. However, more and more local maxima need to be squeezed into the same space. A consequence is that the maxima must be on smaller and smaller scales, and on these scales the landscape becomes more and more similar to the landscape on a larger scale. This is known as scale--invariance or self--similarity and a main prerequisite of fractality~\cite{man83}. 

Similarly to defining ruggedness, also the definition of a fractal is delicate, see for instance the discussion in~\cite{falc90}, p. xviii--xxii.  
This paper subscribes to the definition that a fractal is a set for which its Hausdorff--Besicovitch dimension is larger than its topological dimension. Accordingly, it can be defined that a fitness landscape is fractal if the  Hausdorff--Besicovitch dimension of the fitness distribution over the configuration space is larger than the topological dimension of the configuration space. However, while the  Hausdorff--Besicovitch dimension is suitable for defining the geometric complexity of a fractal, it is very hard to calculate numerically. In Sec. \ref{sec:exp} reporting numerical experiments with dynamic targeting landscapes  (\ref{eq:opttargetdyn}), a widely used technique, box--counting, is employed to characterize the fractality of these landscapes.
The box--counting dimension $D_{BC}$ 
can be rather easily calculated~\cite{block90,molt93} and approximates the Hausdorff--Besicovitch dimension, see  Appendix A. 

A main property of fractal landscapes is that local maxima occur on varying scales of the configuration space variables, which suggests several conclusions. A first is that landscape measures numerically evaluating ruggedness should be scale--invariant and self--similar. For the targeting landscape  (\ref{eq:opttargetdyn}) this means that by varying the scale of the parameters in $\mathcal{P}$,  similar values of  ruggedness measures should be obtained.   A second is that the value of the ruggedness measures should be in relation to the fractal dimension of the landscape.  In other words, the fractal dimension should be a characterization of the degree of ruggedness occurring in the landscape. While fractality can be seen as the upper limit of (scale--variant) ruggedness, the fractal dimension becomes a measure of scale--invariant ruggedness.   
The validity of these conclusions can be tested by numerical experiments, which are also reported in  Sec. \ref{sec:exp}. 
 Therefore, this paper considers two numerical ruggedness measures, correlation length $\lambda$~\cite{wein90,stad96} and information content $h_{IC}$~\cite{mun15,vassi00} to compare ruggedness of dynamic fitness landscapes of the targeting problem, see Appendix B for details.

\section{Numerical experiments} \label{sec:exp}
\begin{figure*}[t]

\includegraphics[trim = 30mm 90mm 30mm 90mm,clip, width=7cm, height=5cm]{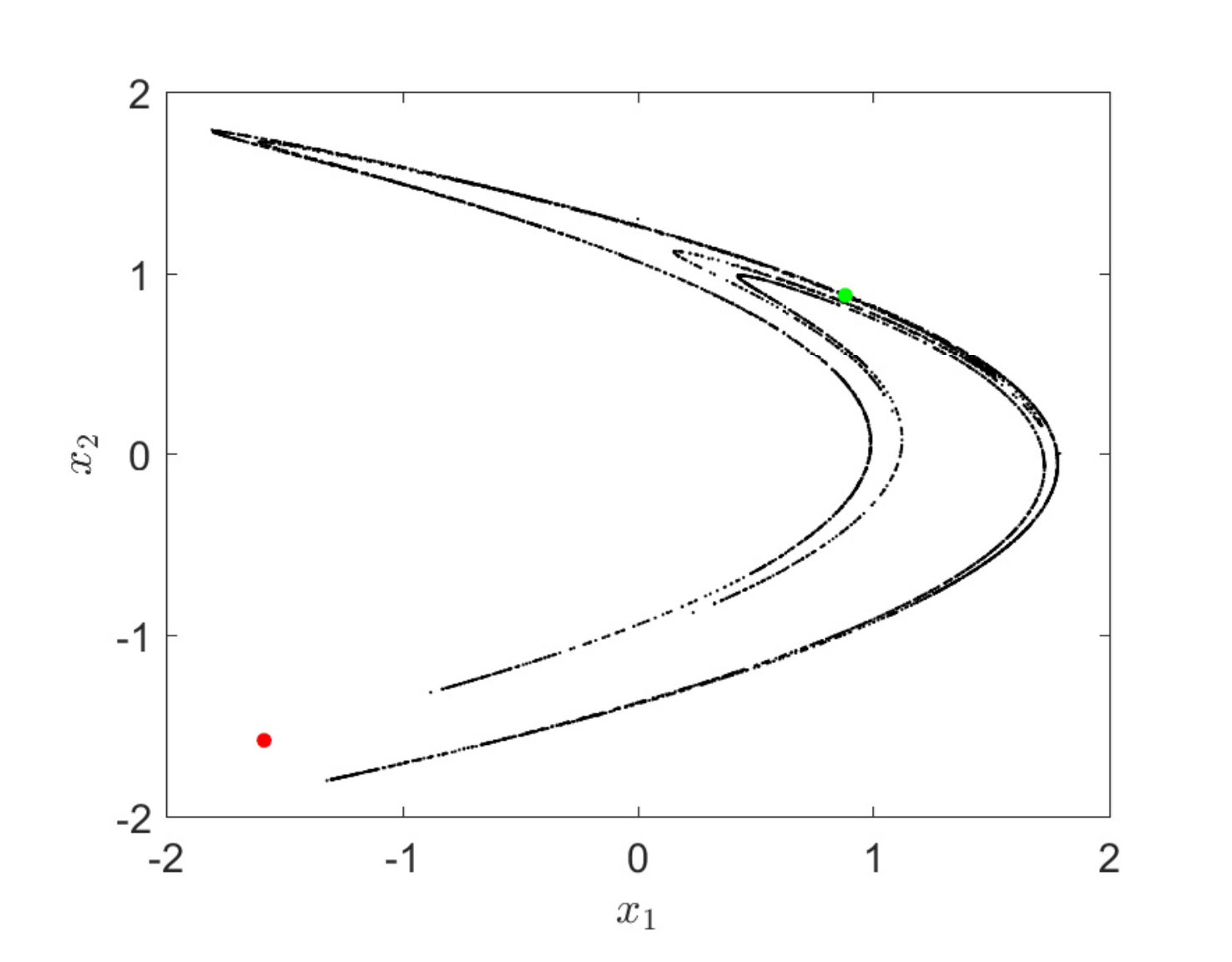} 
\includegraphics[trim = 30mm 90mm 30mm 90mm,clip, width=7cm, height=5cm]{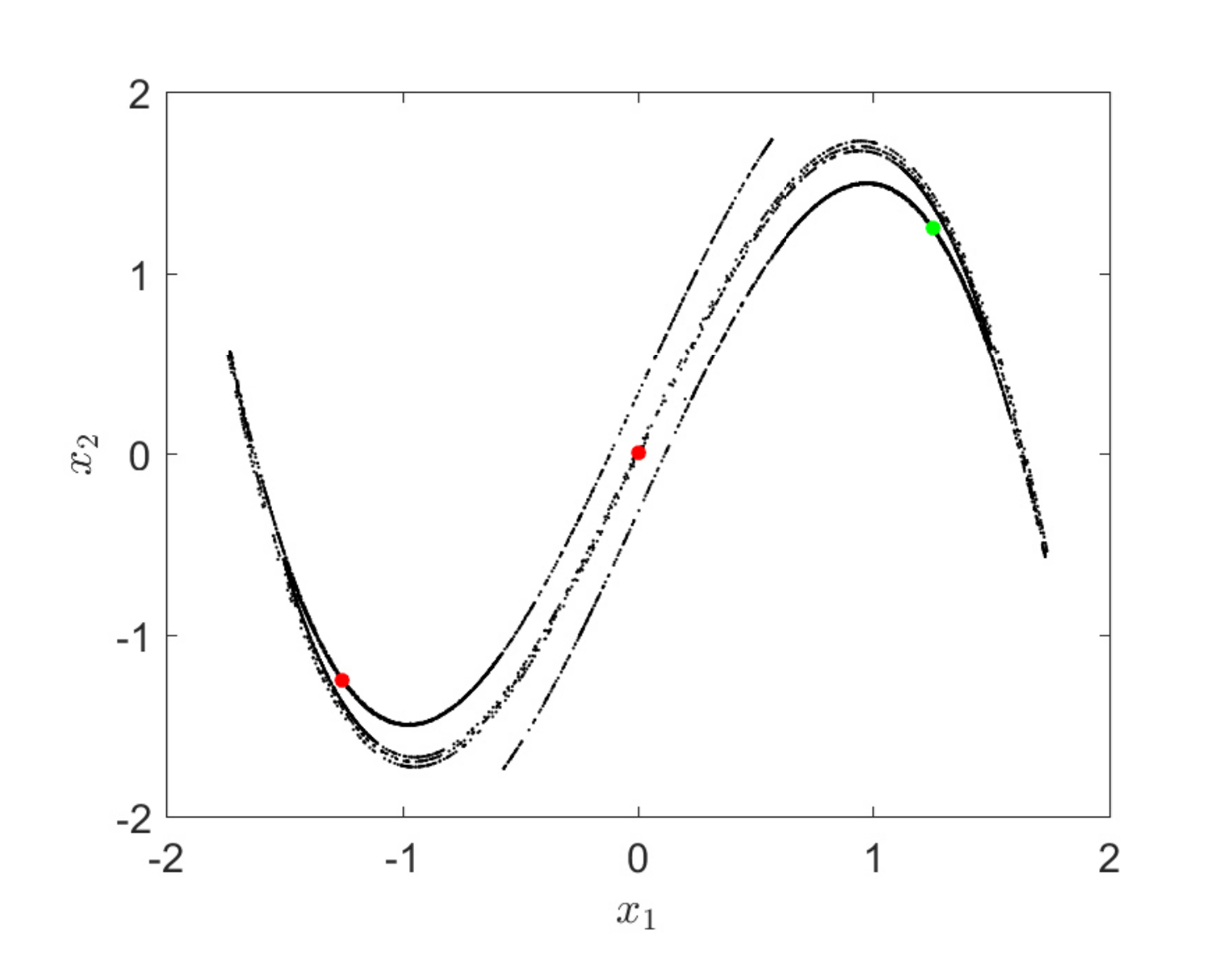} 

\hspace{0.8cm} (a)  \hspace{6.3cm} (b)

\includegraphics[trim = 30mm 90mm 30mm 90mm,clip, width=7cm, height=5cm]{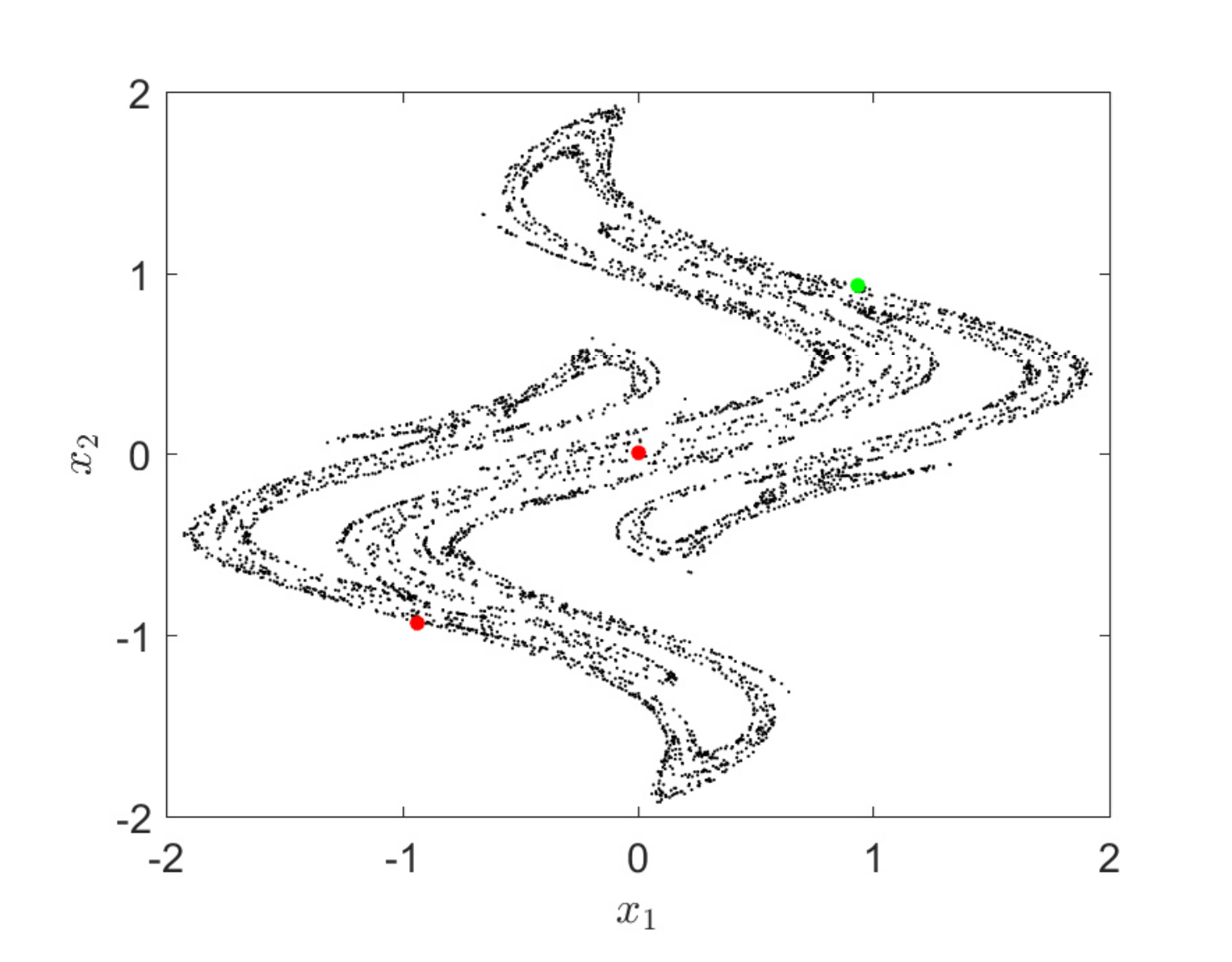} 
\includegraphics[trim = 30mm 90mm 30mm 90mm,clip, width=7cm, height=5cm] {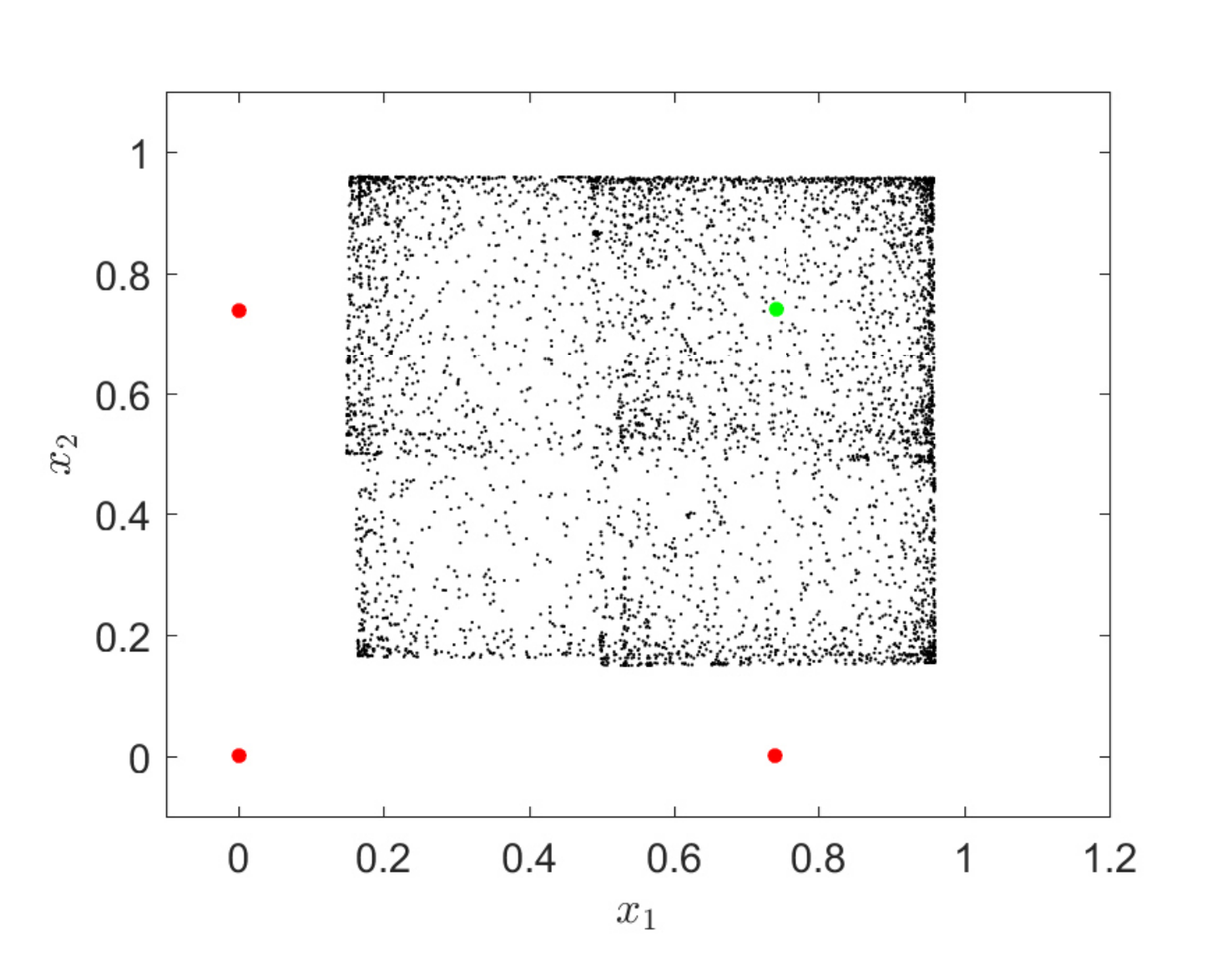} 

\hspace{0.8cm} (c)  \hspace{6.3cm} (d)

\caption{Chaotic attractors, fixed points (depicted as red dots) and target points (depicted as green dots). (a) H\'enon map (\ref{eq:henon}),  (b) Holmes map (\ref{eq:holmes}),  (c) exponential map (\ref{eq:genexp2}), (d) coupled logistic map (\ref{eq:couplog}).   }
\label{fig:1}

\end{figure*}
\subsection{Experimental setup}
The numerical experiments are done with four different dynamical systems: 
the H\'enon map~\cite{hen76} \begin{equation} \label{eq:henon}
x(k+1)=\left( \begin{array}{cc} p_1-x_1(k)^2+p_2x_2(k) \\
x_1(k) \end{array} \right), \end{equation}
the Holmes map~\cite{holm79} \begin{equation} \label{eq:holmes}
x(k+1)=\left( \begin{array}{cc} x_2(k) \\
-p_1x_1(k)+p_2x_2(k)-x_2(k)^3 \end{array} \right), \end{equation}
a 2D exponential map~\cite{ding96,richt08}
\begin{equation}
x(k+1)
=\left( \begin{array} {cc} p_1 x_1(k)  (1-x_1(k)^2) \exp{\left(
-x_1(k)^2 \right)}+p_2
x_2(k)\\
x_{1}(k)\end{array} \right),  \label{eq:genexp2}
\end{equation}
and the coupled logistic map with bilinear coupling~\cite{lloyd95}
\begin{equation}
x(k+1)
=\left( \begin{array} {cc} p_1 x_1(k)  (1-x_1(k))+p_2x_1(k)x_2(k)\\
p_1 x_2(k)  (1-x_2(k))+p_2x_1(k)x_2(k)\end{array} \right).  \label{eq:couplog}
\end{equation}
Each map depends on two parameters $(p_1,p_2)$ to be used as control perturbations for targeting, which means that the configuration space (\ref{eq:limit}) is 2D for the examples. According to the general description (\ref{eq:basic}), the maps (\ref{eq:henon})--(\ref{eq:couplog}) also depend on the time variable $k$.  Hence, the dynamic fitness landscapes (\ref{eq:opttargetdyn}) can be calculated over the parameter space  $\mathcal{P}=\{p\in \mathbb{R}^2 | \|p-\bar{p} \|_\infty \leq \eta \}$, for each $k=1,2,\ldots, \kappa$ and any given $\eta$. In other words, the dynamic fitness landscapes depends implicitly on the time variable $k$ of the dynamical systems as they are calculated for a target time $\kappa$ that can have any value of $k>1$. 

The systems (\ref{eq:henon})--(\ref{eq:couplog})  all show chaotic behavior,  but have slightly different chaoticity and attractor dimension, see Tab. \ref{tab:1} for parameters $\bar{p}$,  Lyapunov exponents $(h_1,h_2)$, Kaplan--Yorke dimensions $D_{KY}$~\cite{kap79,rich08a}, and box--counting dimensions $D_{BC}$ of the attractors.The Lyapunov exponents are calculated using QR--factorization~\cite{vonbre97},
which is considered to be computationally efficient, reliable and robust. The Kaplan--Yorke dimensions for 2D systems with $h_1>0$ and $h_2<0$ is $D_{KY}=1+h_1/|h_2|$, see~\cite{kap79,rich08a}. The box--counting dimensions are computed considering the chaotic attractor of each system, see Fig. \ref{fig:1}, and employing the algorithm given in Appendix A with $j=6$ and $\delta_0=1/2024$.  All largest Lyapunov exponents are positive, indicating that the dynamics is chaotic. The chaotic attractors have all  non--integer dimensions as expressed by both $D_{KY}$ and $D_{BC}$.  It can further be seen that for the maps (\ref{eq:henon})--(\ref{eq:genexp2}) both fractal dimensions $D_{KY}$ and $D_{BC}$ give very similar results. 

 The coupled logistic map (\ref{eq:couplog}) has two positive Lyapunov exponents for the parameter values considered, and hence differs from the other three systems. It is hyperchaotic, which may occur in 2D maps that are not invertible, which applies to Eq.  (\ref{eq:couplog}). The dynamical system  (\ref{eq:couplog}) is furthermore not an area--contracting map and forms a chaotic attractor with a fractal dimension close to $2$. By definition, the Kaplan--Yorke dimension of this chaotic attractor is equal to 2, which is not meaningful because it does not reflect its fractal characteristics. The result for the box--counting dimension $D_{BC}$ of this attractor is a good approximation.

As target points one of the systems' fixed points are used, namely for the H\'enon map
\begin{equation*}\bar{x}=\frac{1}{2}\left(-(1-\bar{p}_2)+\sqrt{(1-\bar{p}_2)^2+4\bar{p}_1}\right)  \: \cdot \: \left(1,1 \right)^T\end{equation*}
 for the Holmes map
\begin{equation*}\bar{x}= \sqrt{\bar{p}_2-\bar{p}_1-1}  \: \cdot \: \left(1,1 \right)^T, \end{equation*} 
for the exponential map
\begin{equation*}\bar{x}=\sqrt{1-\mathcal{W}\left(\frac{1-\bar p_2}{\bar p_1}\exp{(1)}\right)} \: \cdot \: \left(1,1 \right)^T, \end{equation*} with $\mathcal{W}\left(\frac{1-\bar p_2}{\bar p_1}\exp{(1)}\right)$ the Lambert
$\mathcal{W}$ function (see e.g. ~\cite{corless96,vall00}) of
$\frac{1-\bar p_2}{\bar p_1}\exp{(1)}$,
and for the coupled logistic map 
\begin{equation*}\bar{x}=\frac{\bar p_1(\bar p_1+\bar p_2-1)-\bar p_2}{\bar p_1^2-\bar p_2^2} \: \cdot \: \left(1,1 \right)^T .  \end{equation*}  Fig. \ref{fig:1} shows the chaotic attractors of these four example systems with the fixed points used as targets indicated as green dots. Note that only fixed points embedded in the attractor can be taken as target points as only for these points orbit density and ergodicity is given. 
\begin{table}
\caption{Parameter values $p(k)=\left(\bar{p}_1,\bar{p}_2 \right)$,  Lyapunov exponents $(h_1,h_2)$, Kaplan--Yorke dimension $D_{KY}$ and box--counting dimension $D_{BC}$ of the dynamical systems used as examples.}
\label{tab:1}
\begin{tabular}{|c|c|c|c|c|}
System &  $p(k)=\left(\bar{p}_1,\bar{p}_2 \right)$ &  $(h_1,h_2)$ & $D_{KY}$ & $ D_{BC}$  \\ \hline
 (\ref{eq:henon})  & $(1.40,0.30)$ & $(0.4198,-1.6238)$ & $1.2586$ & $1.2738 \pm 0.3419$ \\ \hline (\ref{eq:holmes})  & $(0.20,2.77)$ & $(0.5975, -2.2069)$ & $1.2707$ & $1.2886 \pm 0.3382$ \\ \hline
(\ref{eq:genexp2}) & $(3.50,0.81)$ & $(0.4406,-0.6513)$ & $1.6765$ & $1.6325 \pm 0.1948$ \\ \hline 
(\ref{eq:couplog}) & $(3.82,0.01)$ &  (0.4055,0.2899) & -- & $1.9004 \pm 0.2262$ \\ \hline

\end{tabular}

\end{table}

The calculation of the box--counting dimension $D_{BC}$ has been carried out with $j=6$ repetition starting from an initial edge length $\delta_0$. The effect of $\delta_0$ on calculating $D_{BC}$ is analyzed by numerical experiments reported in the next section. Computing the landscape measures correlation length $\lambda$ and information content $h_{IC}$ has been done with a random walk of length $T=17500$ and $\epsilon=0.25$. The results are averaged over $150$ repetitions from different initial points of the walks. Experiments have shown that the results are statistically equivalent for different initial points, which makes it reasonable to assume that the targeting landscapes considered are isotropic.  

\begin{figure*}[tb]

\includegraphics[trim = 30mm 90mm 30mm 90mm,clip, width=7cm, height=5cm]{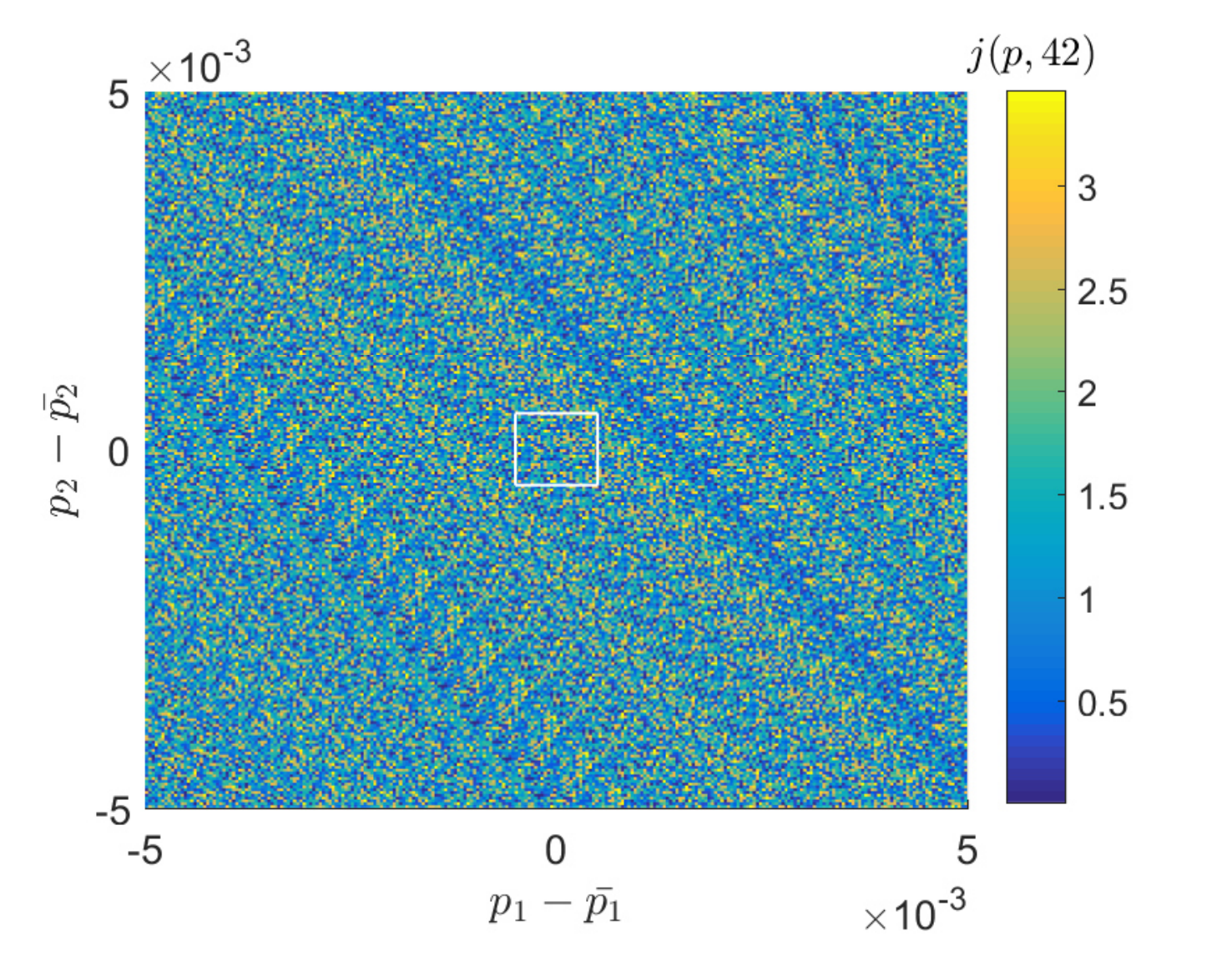} 
\includegraphics[trim = 30mm 90mm 30mm 90mm,clip, width=7cm, height=5cm]{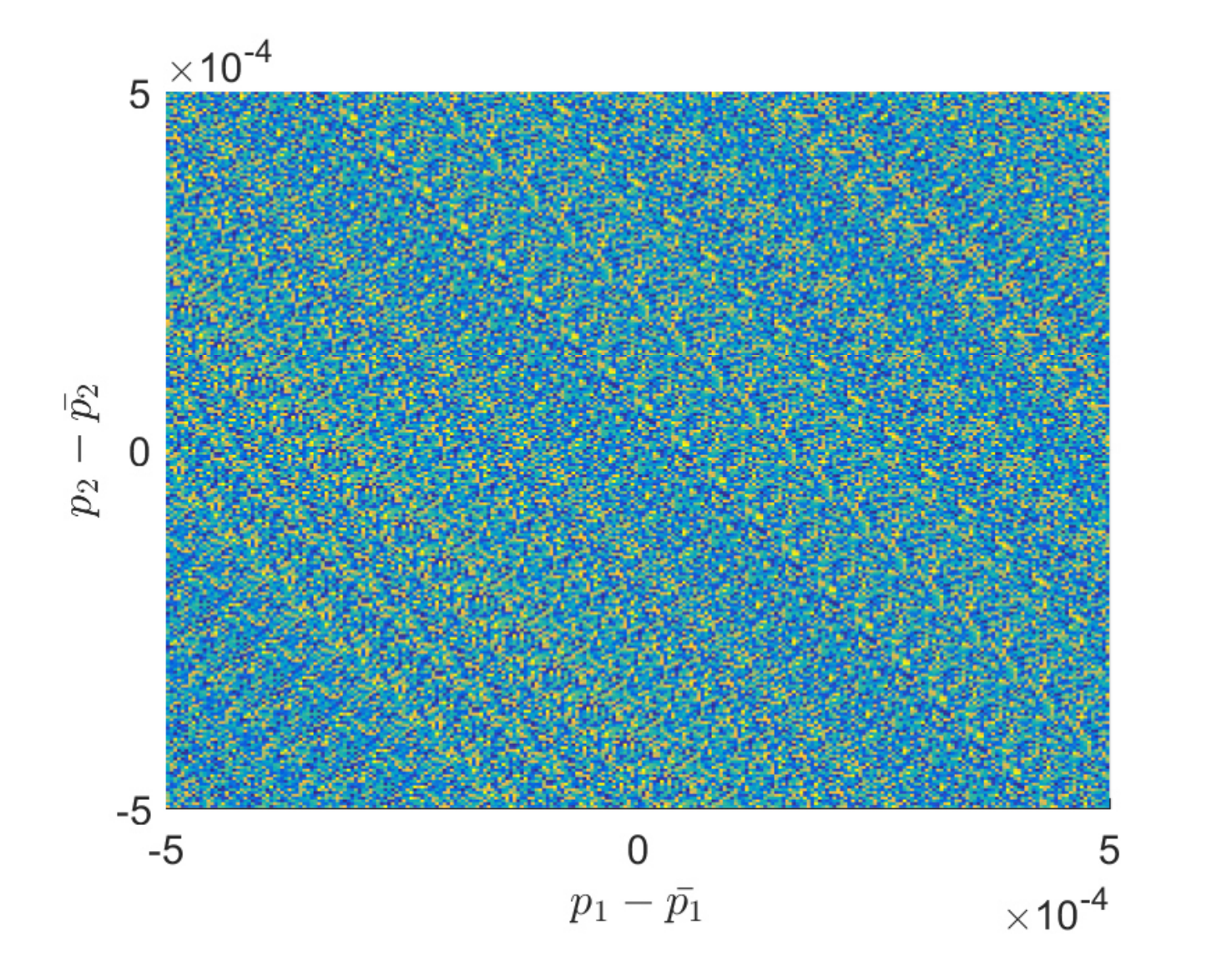} 

\hspace{0.8cm} (a)  \hspace{6.3cm} (b)

\includegraphics[trim = 30mm 90mm 30mm 90mm,clip, width=7cm, height=5cm]{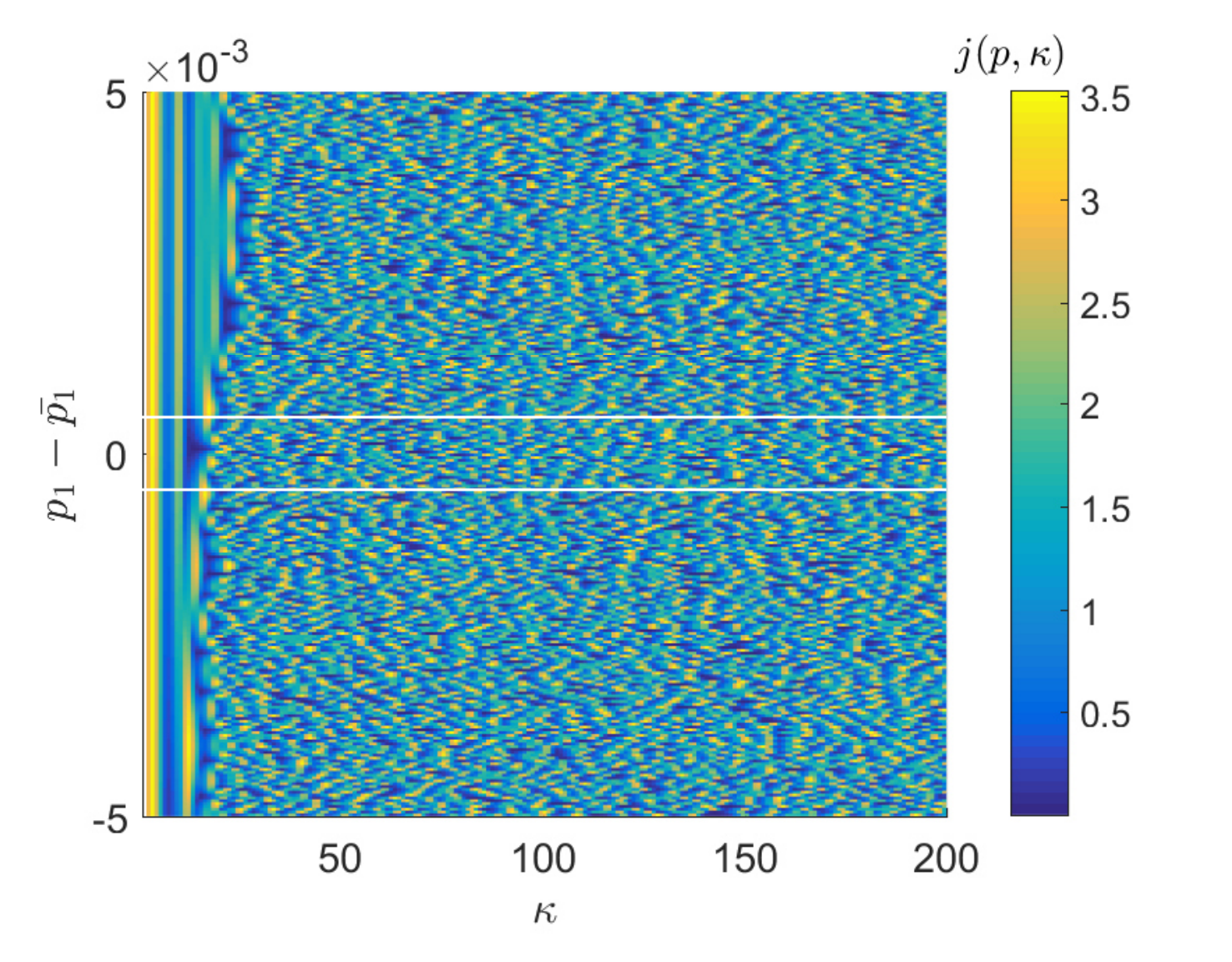} 
\includegraphics[trim = 30mm 90mm 30mm 90mm,clip, width=7cm, height=5cm] {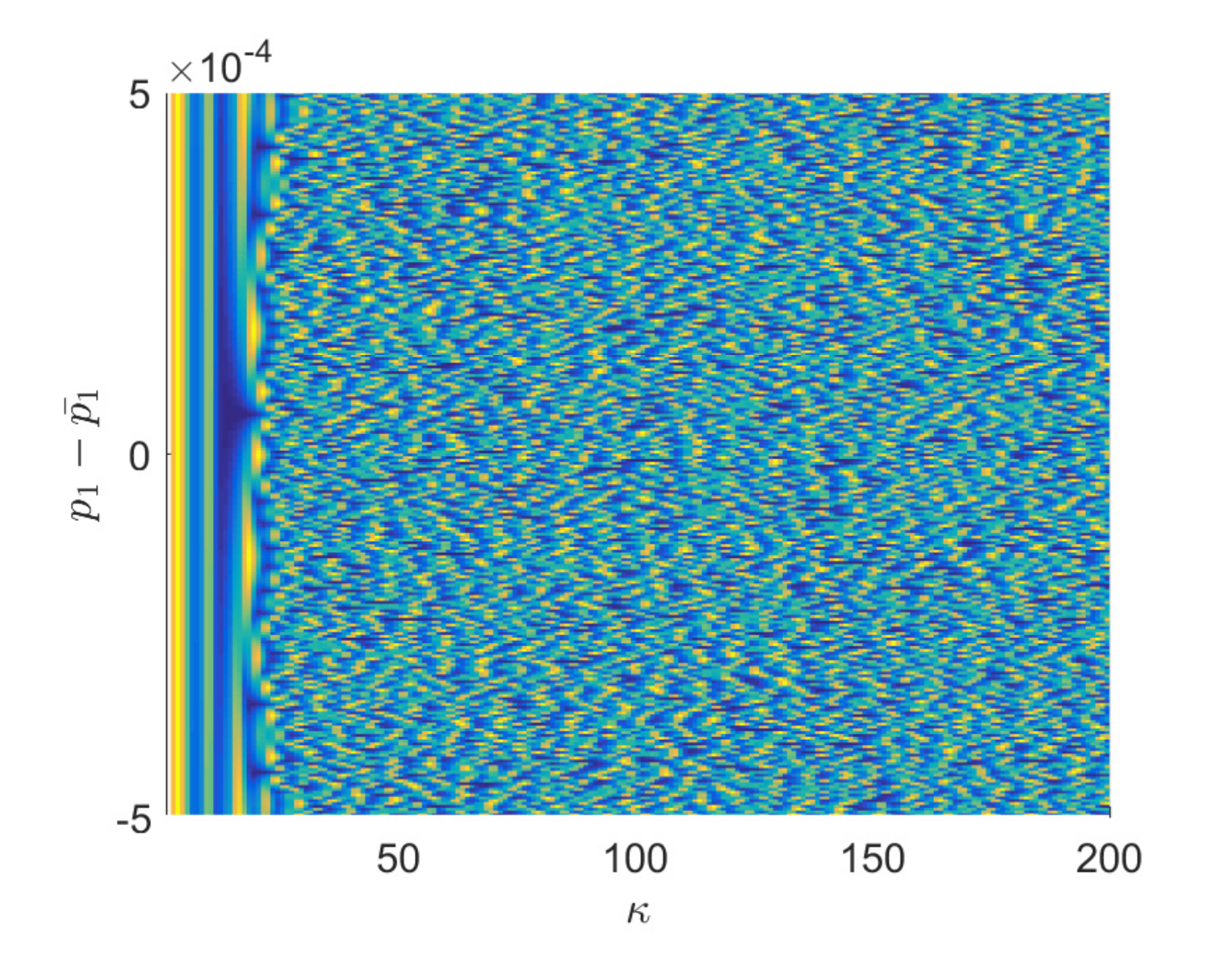} 

\hspace{0.8cm} (c)  \hspace{6.3cm} (d)

\caption{Fractal landscapes in targeting problems of the H\'enon map (\ref{eq:henon}).  The dynamic distance function (\ref{eq:opttargetdyn}) is shown over parameter perturbation $p=(p_1-\bar p_1,p_2-\bar p_2)$ for target time $\kappa=42$: (a) $\eta=0.01$, (b) magnification of the area withn the white lines of (a) with $\eta=0.001$.   Variable target time $\kappa$ over parameter perturbation $p=(p_1-\bar p_1,\bar p_2)$: (c) $\eta=0.01$, (d) magnification of the area within the white lines of (c) with $\eta=0.001$. }
\label{fig:1a}

\end{figure*}
\subsection{Experimental results and discussion}
In the following, ruggedness and fractality of the dynamic fitness landscapes (\ref{eq:opttargetdyn}) are analyzed by numerical experiments. To illustrate by a graphic example, Fig. \ref{fig:1a} shows the fractal landscape of the targeting problem of the  H\'enon map (\ref{eq:henon}) with the parameter values given in Tab. \ref{tab:1}. The distance function $j(p,\kappa)$ is depicted as a function of $p=(p_1-\bar p_1,p_2-\bar p_2)$ and a constant  $\kappa=42$ in  Fig. \ref{fig:1a}a,b  and as a function of $\kappa$ and $p=(p_1-\bar p_1,\bar p_2)$ (thus for a constant $p_2=0$) in   Fig. \ref{fig:1a}c,d. The Fig.  \ref{fig:1a}b,d are magnifications of the region within the white lines in  Fig.  \ref{fig:1a}a,c. The color bar gives the values of $j(p,\kappa)$. The landscapes are scaled and spatially partitioned by $\eta/256$. Thus, the images of the landscapes in Fig. \ref{fig:1a}a,b contain $256 \times 256$ data points, while in Fig. \ref{fig:1a}c,d, we have $256 \times 200$ data points as the target time is varied for $1 \leq \kappa \leq 200$. The results are typical and can similarly be found for the other maps.
\begin{figure*}[t]

\includegraphics[trim = 30mm 90mm 30mm 90mm,clip, width=7cm, height=5cm]{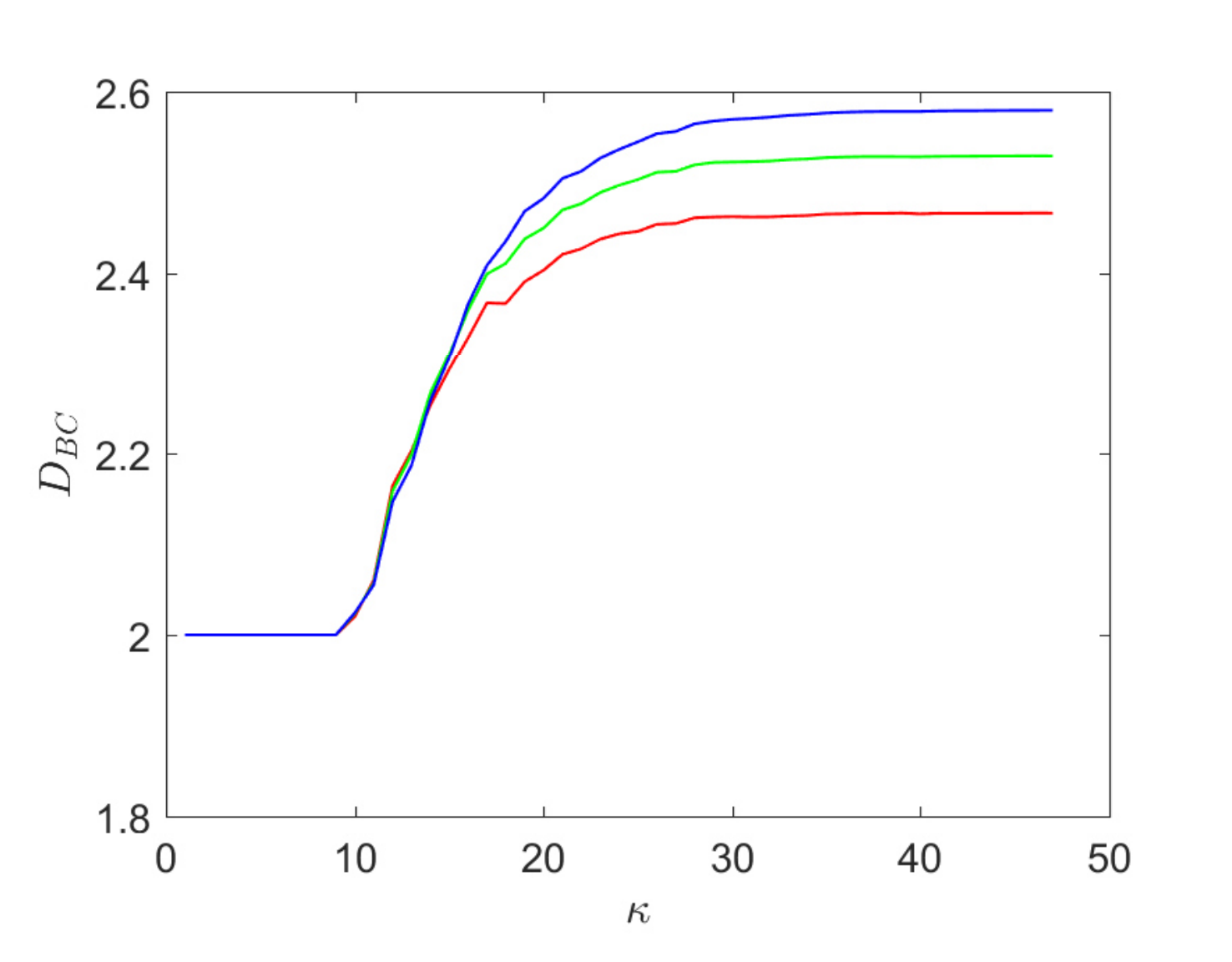} 
\includegraphics[trim = 30mm 90mm 30mm 90mm,clip, width=7cm, height=5cm]{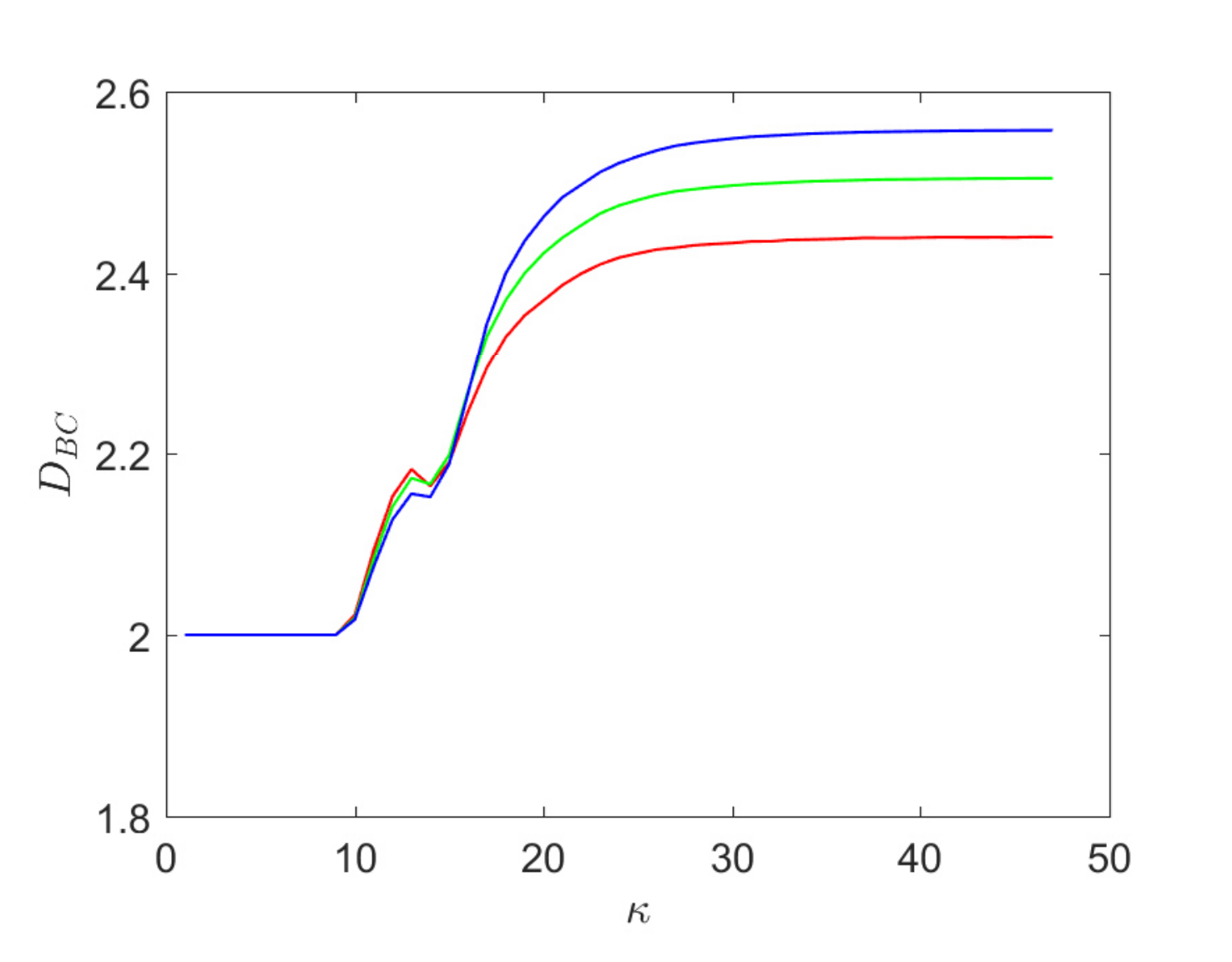} 

\hspace{0.8cm} (a)  \hspace{6.3cm} (b)

\includegraphics[trim = 30mm 90mm 30mm 90mm,clip, width=7cm, height=5cm]{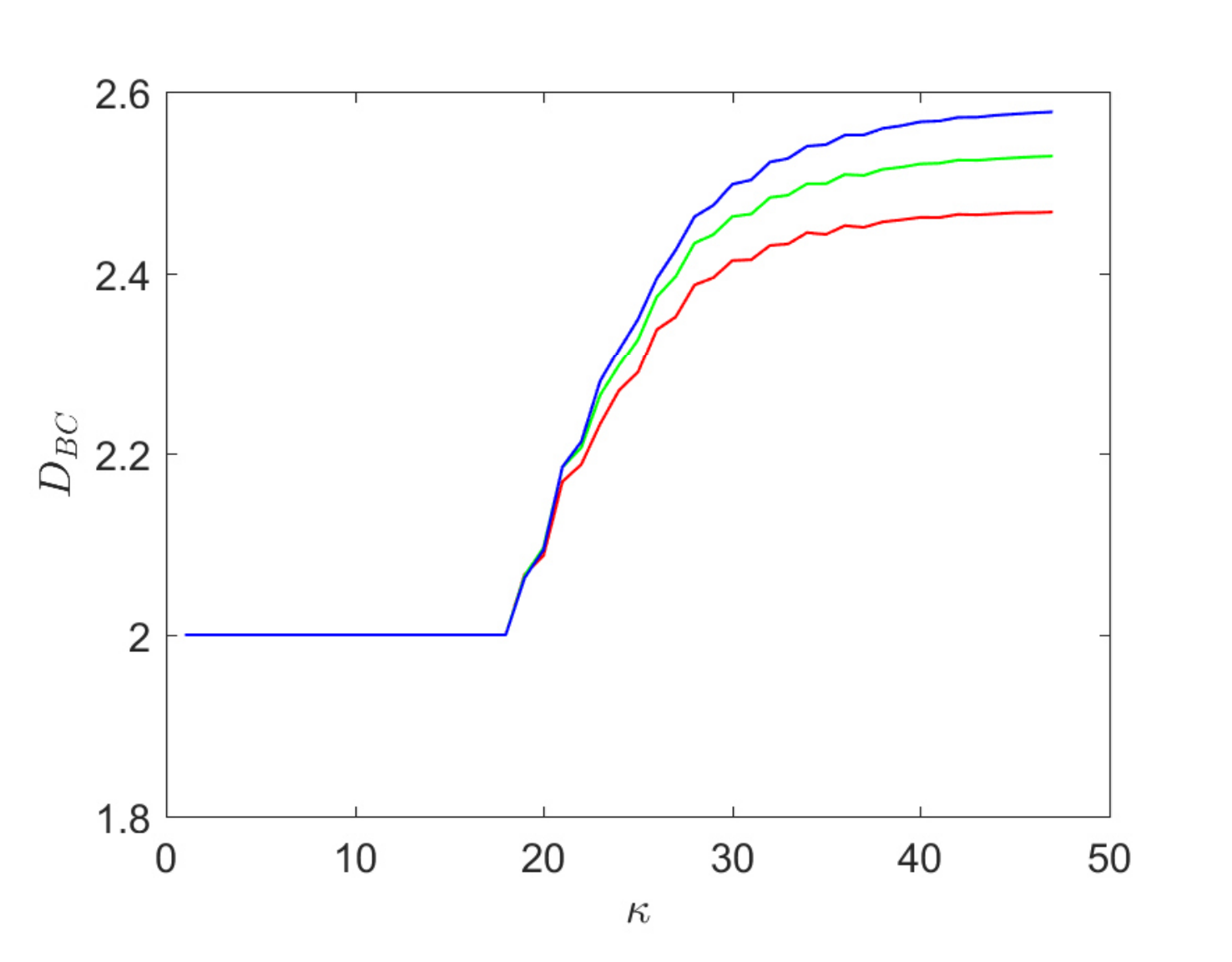} 
\includegraphics[trim = 30mm 90mm 30mm 90mm,clip, width=7cm, height=5cm] {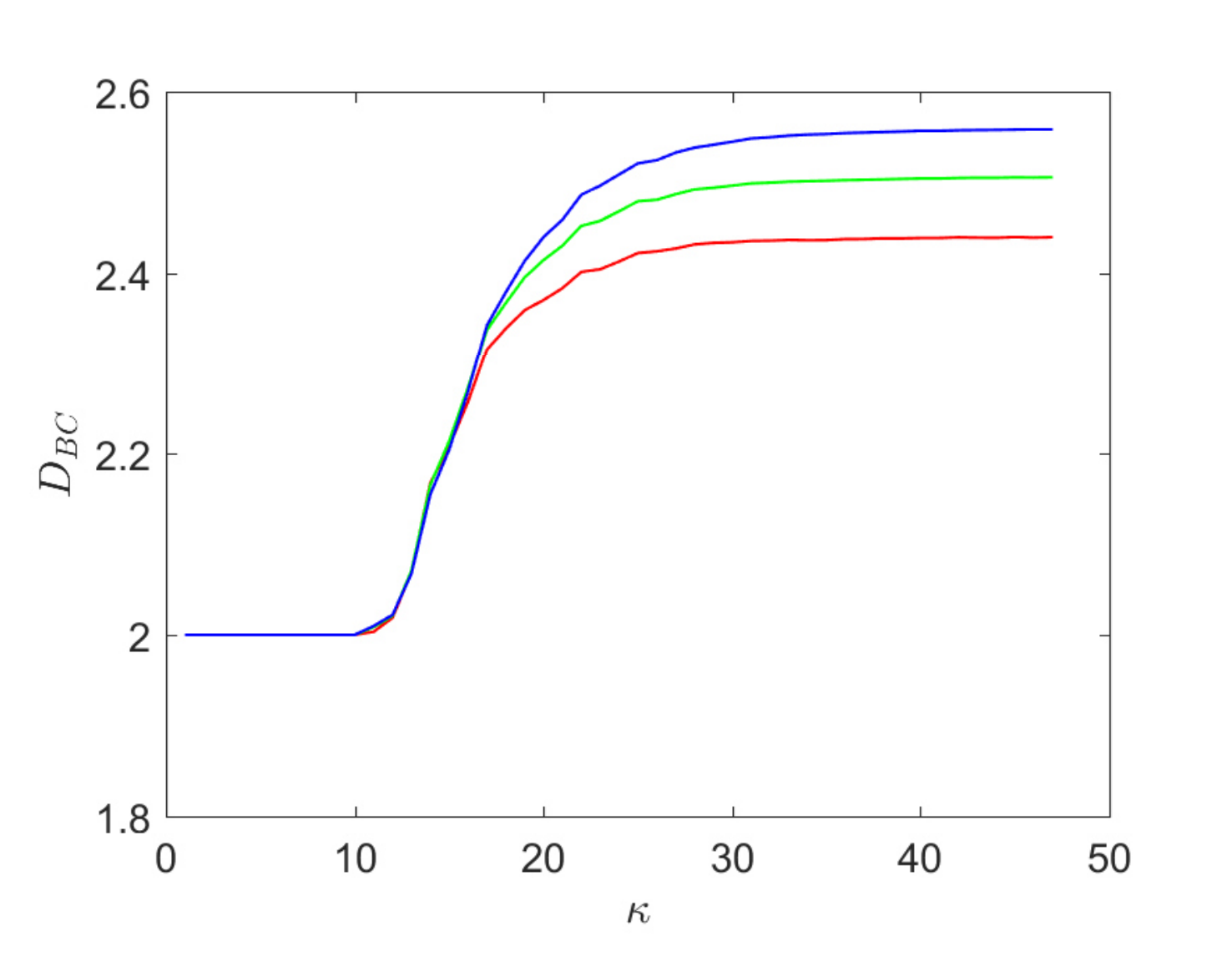} 

\hspace{0.8cm} (c)  \hspace{6.3cm} (d)

\caption{Box--counting dimension $D_{BC}$ over variable target time $\kappa$ for $\eta=0.01$;  $\delta_0=\eta/64$ (red line), $\delta_0=\eta/128$ (green line), $\delta_0=\eta/256$ (blue line).  (a) H\'enon map (\ref{eq:henon}),  (b) Holmes map (\ref{eq:holmes}),  (c) exponential map (\ref{eq:genexp2}), (d) coupled logistic map (\ref{eq:couplog}).    }
\label{fig:2}

\end{figure*}

\begin{figure*}

\includegraphics[trim = 30mm 90mm 30mm 90mm,clip, width=7cm, height=5cm]{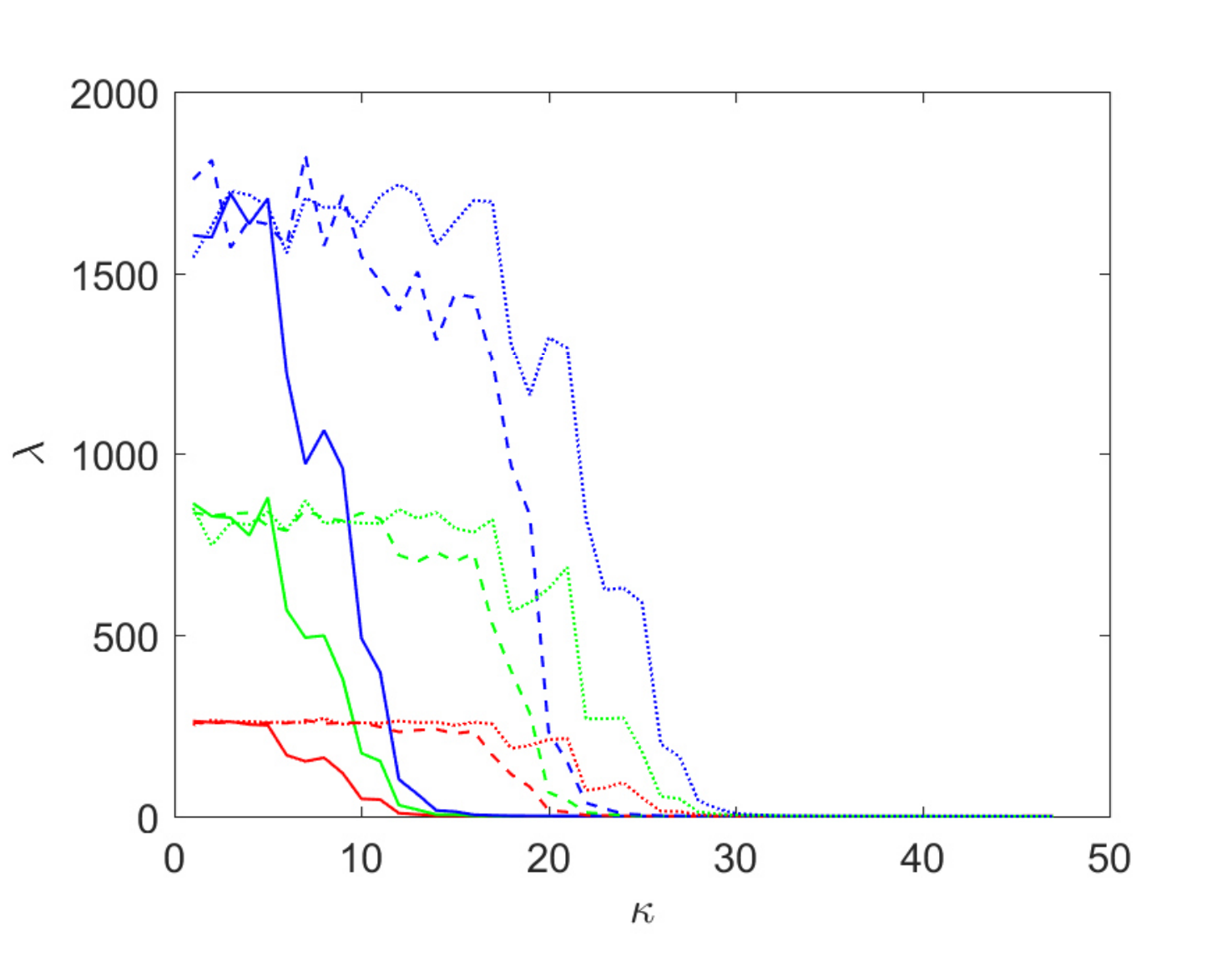} 
\includegraphics[trim = 30mm 90mm 30mm 90mm,clip, width=7cm, height=5cm]{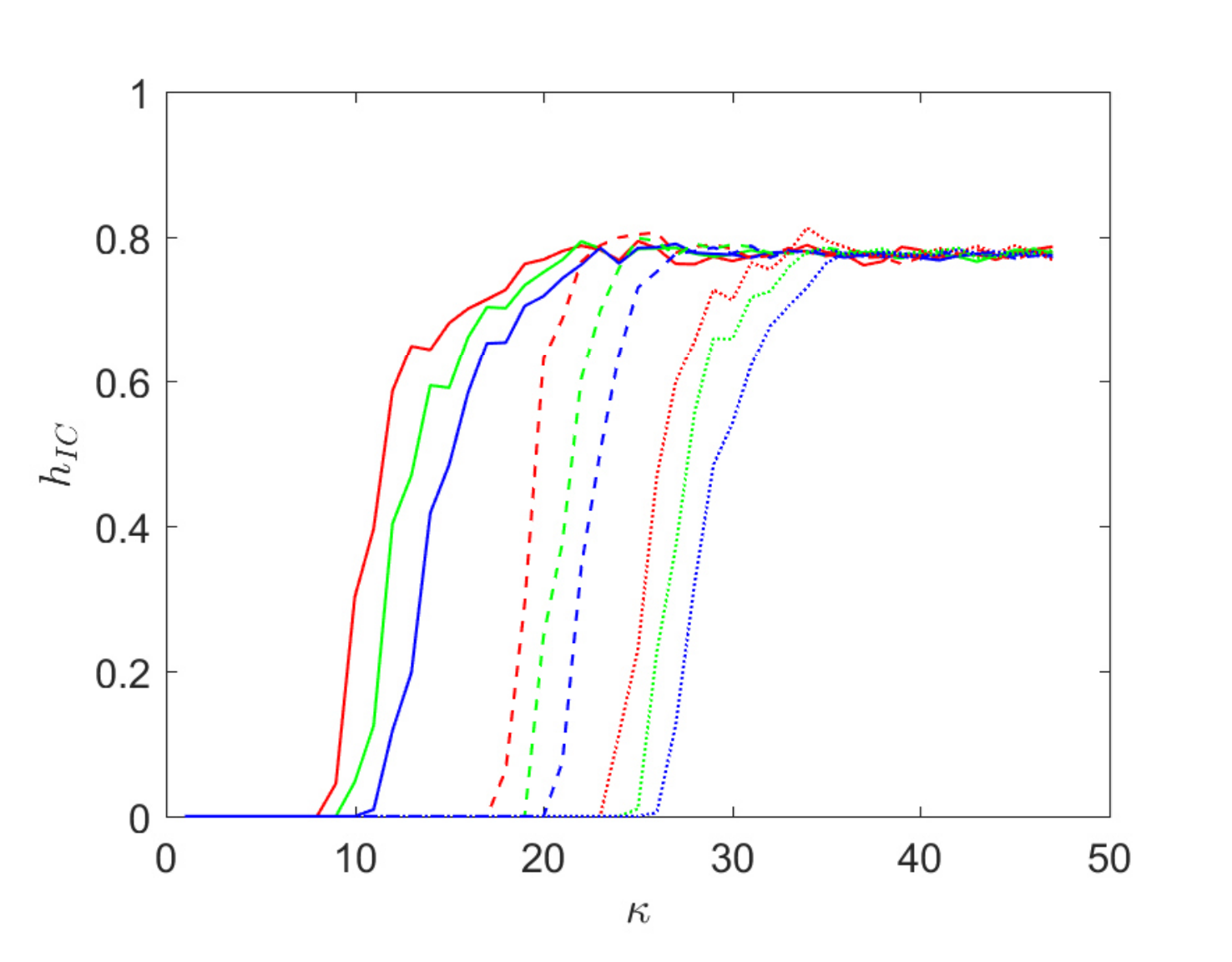} 

\hspace{0.8cm} (a)  \hspace{6.3cm} (b)

\includegraphics[trim = 30mm 90mm 30mm 90mm,clip, width=7cm, height=5cm]{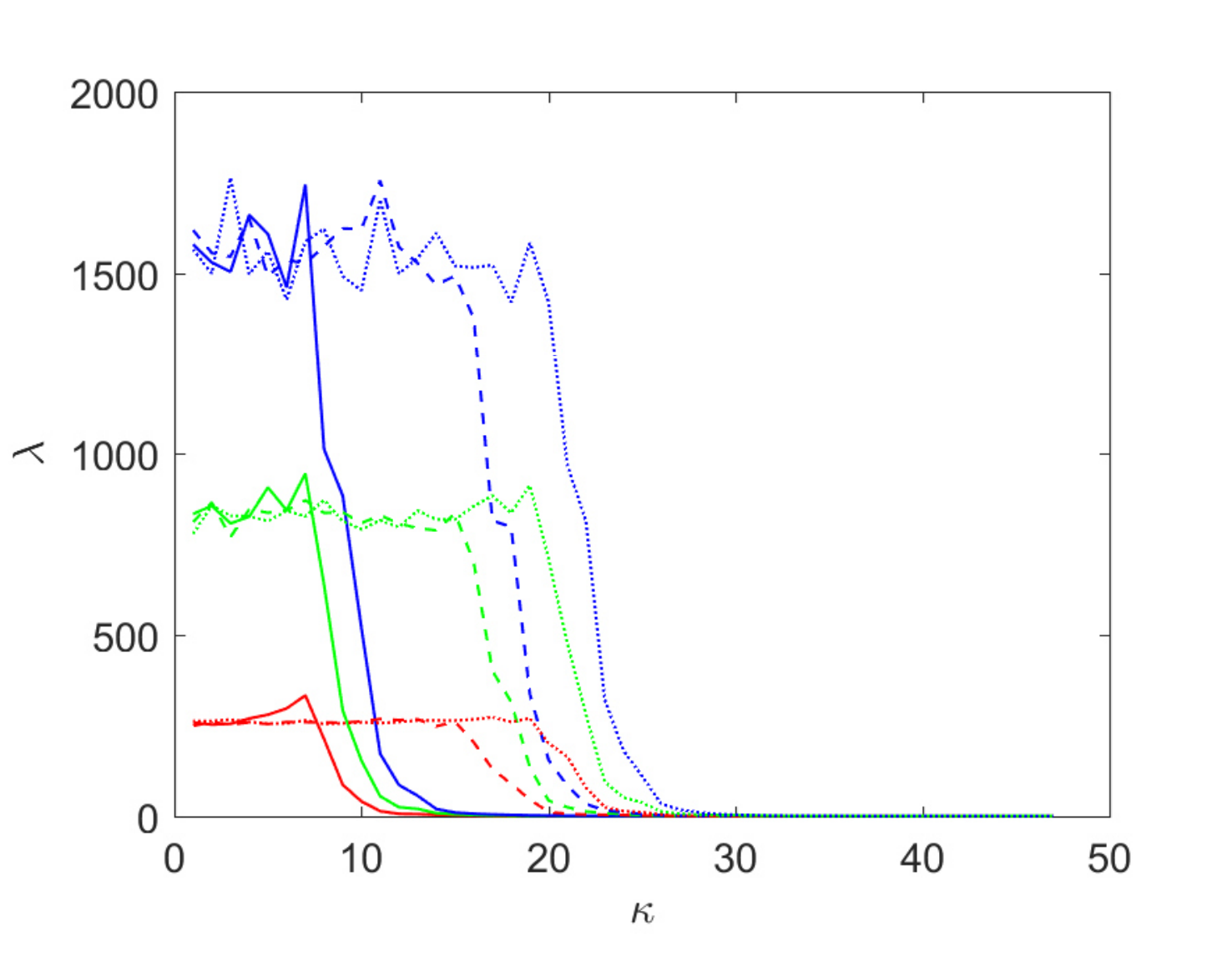} 
\includegraphics[trim = 30mm 90mm 30mm 90mm,clip, width=7cm, height=5cm] {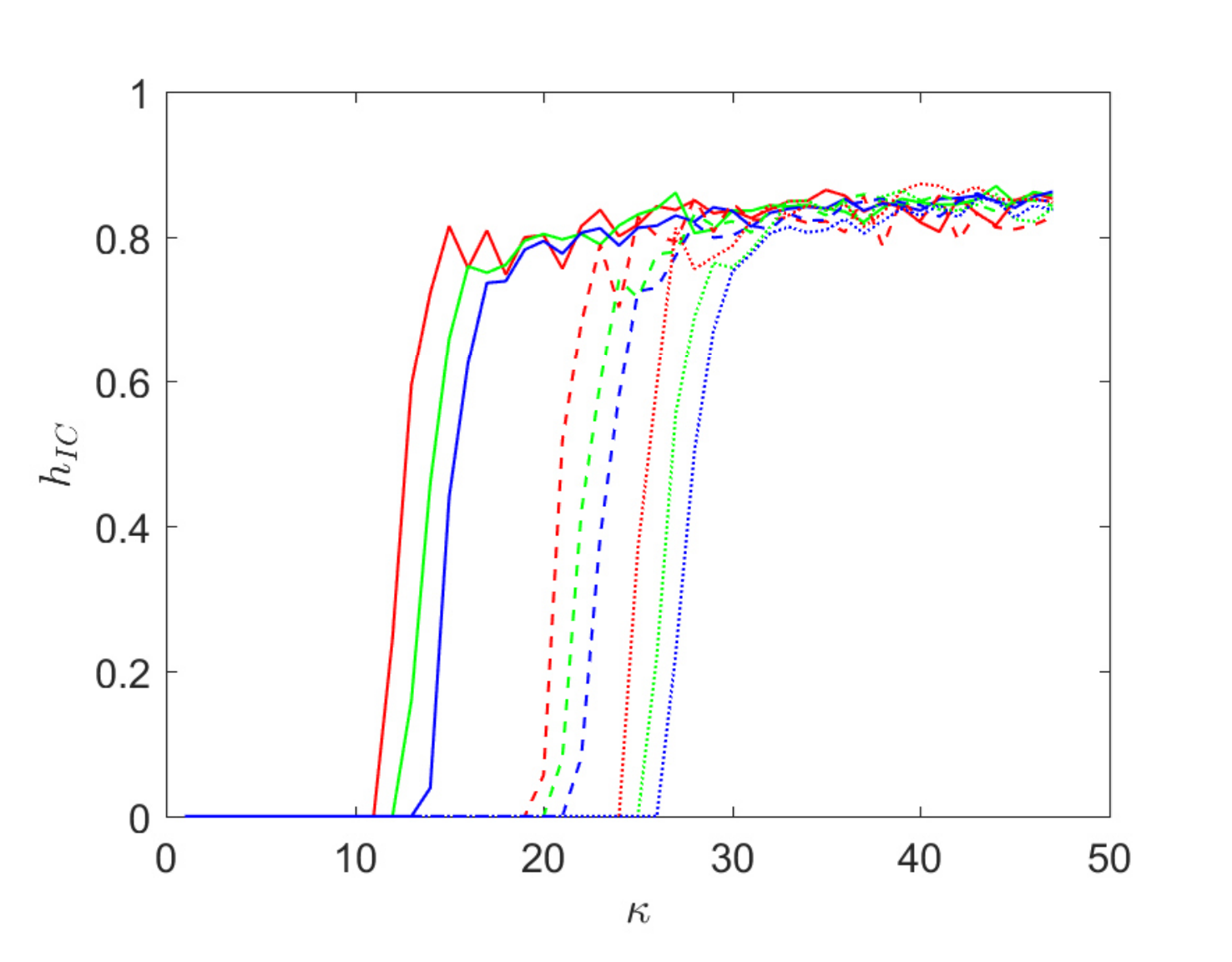} 

\hspace{0.8cm} (c)  \hspace{6.3cm} (d)

\includegraphics[trim = 30mm 90mm 30mm 90mm,clip, width=7cm, height=5cm]{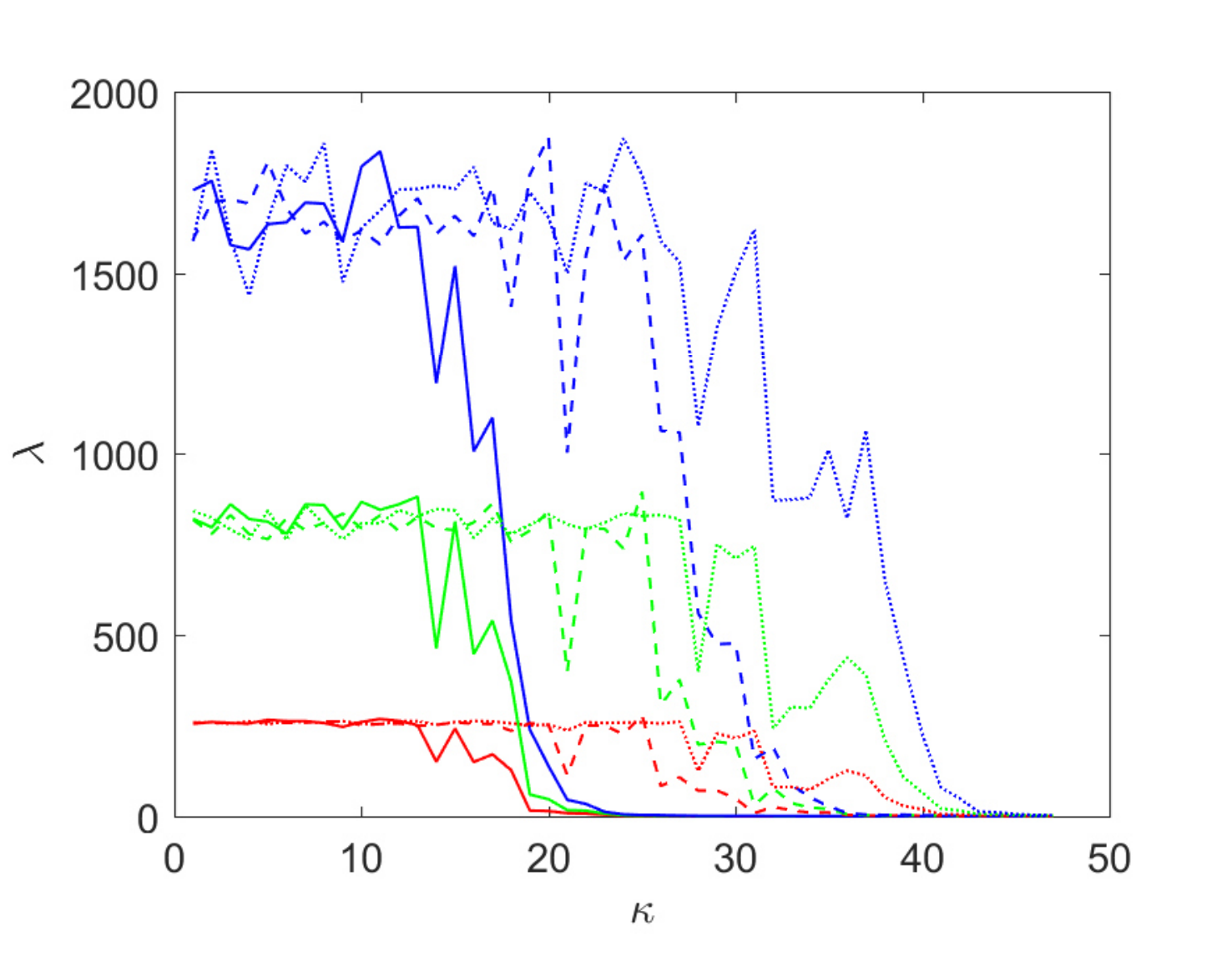} 
\includegraphics[trim = 30mm 90mm 30mm 90mm,clip, width=7cm, height=5cm]{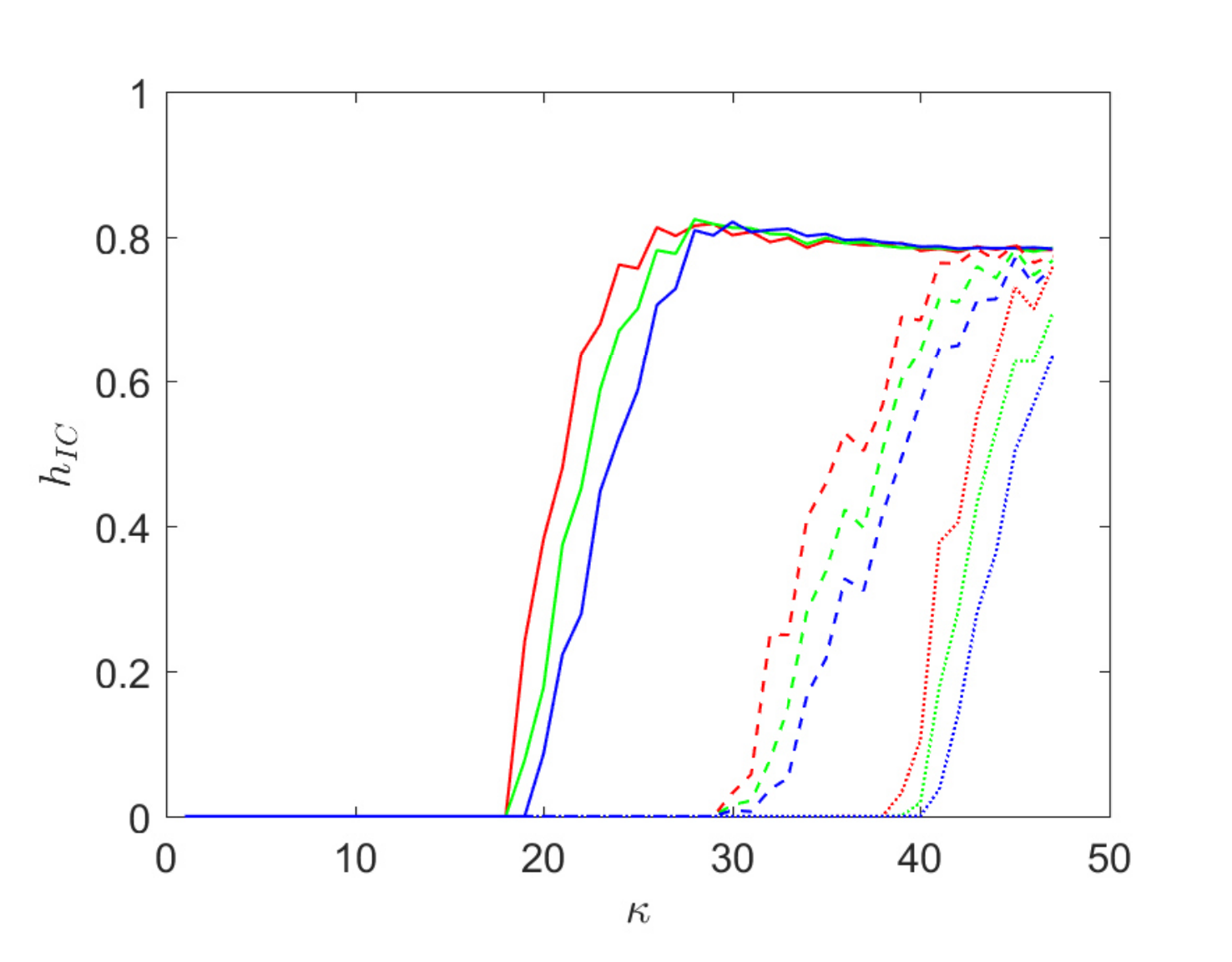} 

\hspace{0.8cm} (e)  \hspace{6.3cm} (f)

\includegraphics[trim = 30mm 90mm 30mm 90mm,clip, width=7cm, height=5cm]{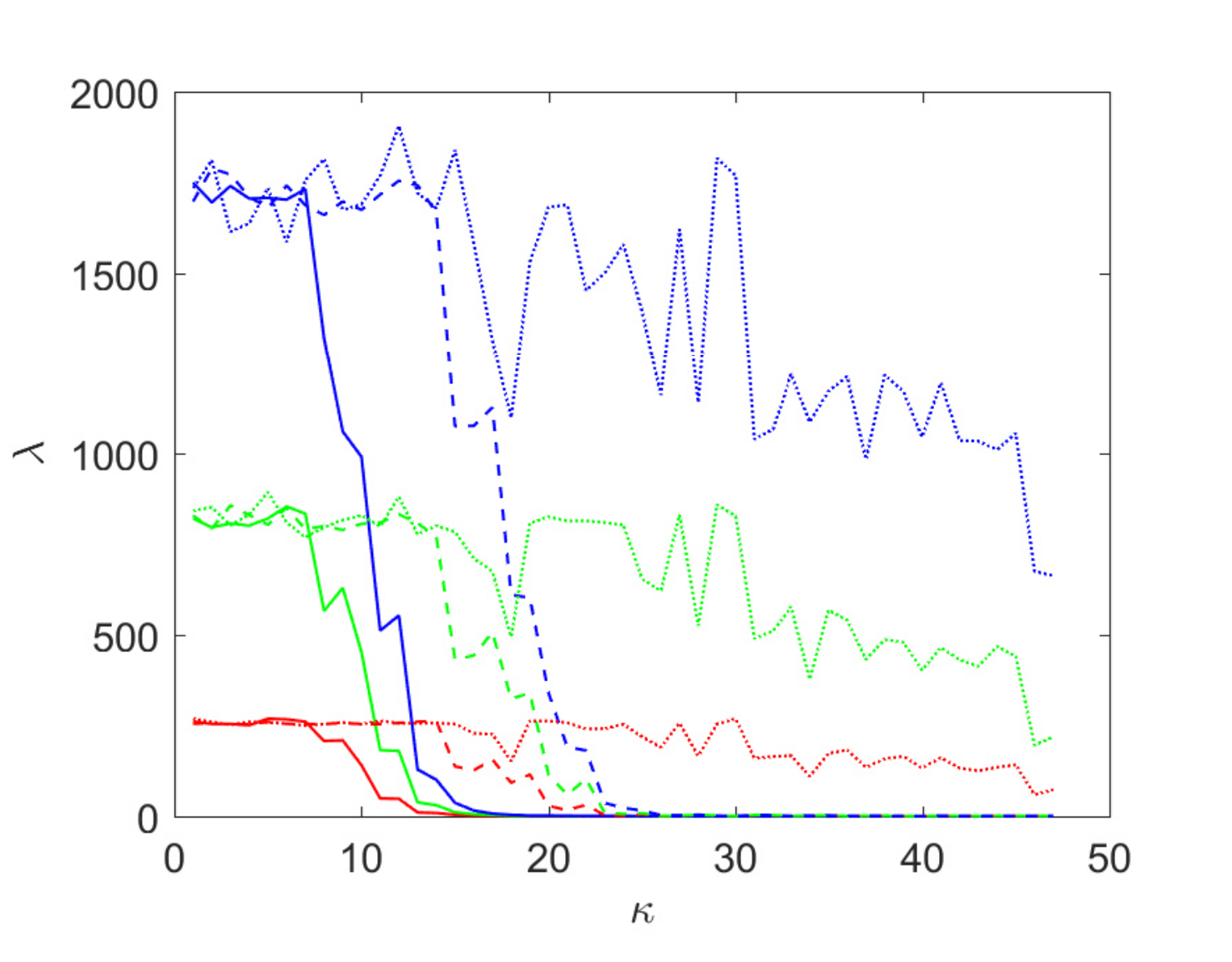} 
\includegraphics[trim = 30mm 90mm 30mm 90mm,clip, width=7cm, height=5cm] {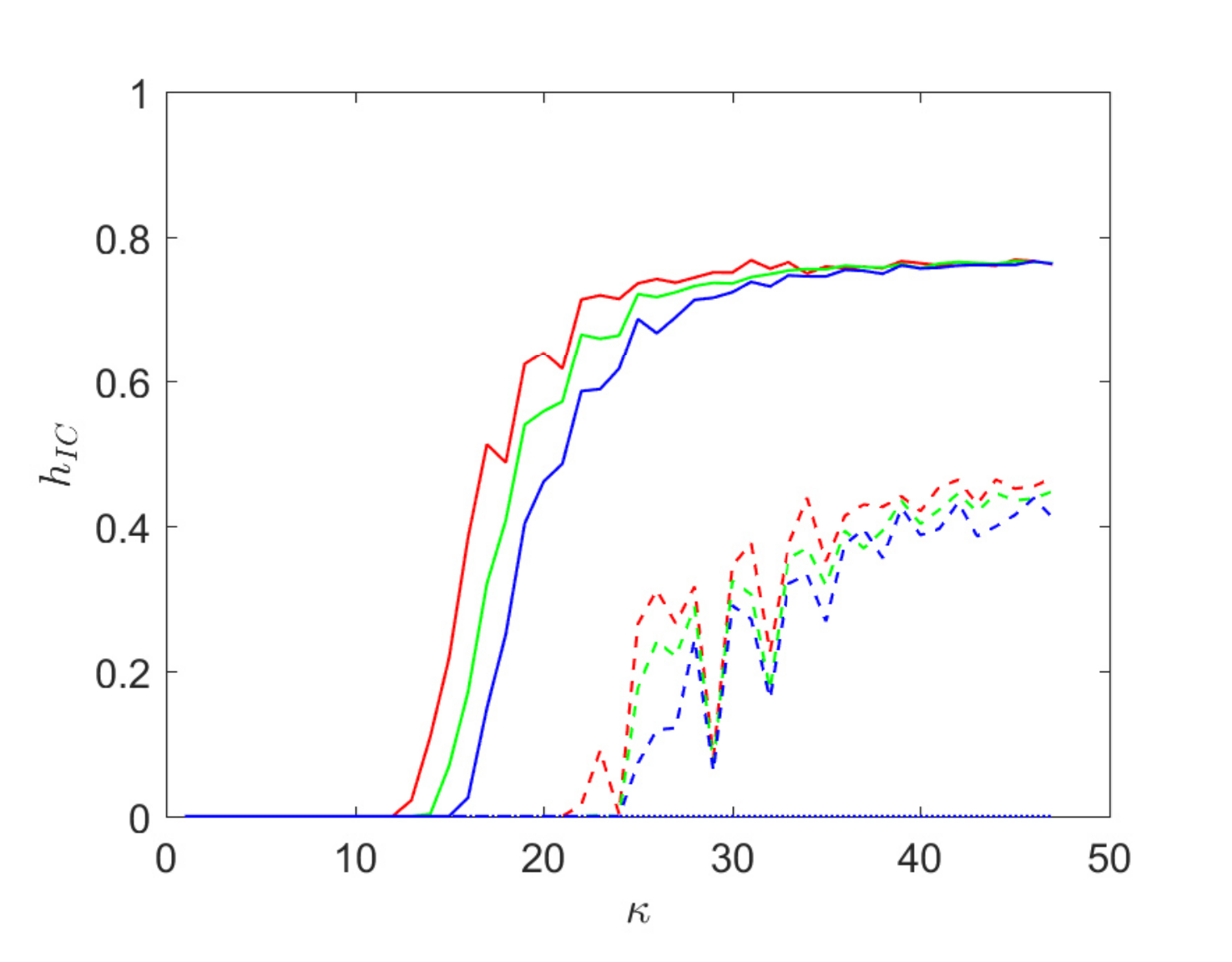} 

\hspace{0.8cm} (g)  \hspace{6.3cm} (h)

\caption{Correlation length $\lambda$ and information content $h_{IC}$ over variable target time $\kappa$. Varying scale $\eta=0.01$ (solid lines), $\eta=0.0001$ (dashdot lines), $\eta=0.000001$ (dotted line);  $\delta_0=\eta/64$ (red lines), $\delta_0=\eta/128$ (green lines), $\delta_0=\eta/256$ (blue lines).  (a), (b) H\'enon map (\ref{eq:henon}),  (c),(d) Holmes map (\ref{eq:holmes}),  (e),(f) exponential map (\ref{eq:genexp2}), (g), (h) coupled logistic map (\ref{eq:couplog}).    }
\label{fig:3}

\end{figure*}

We can see that the landscape in Fig.  \ref{fig:1a}a has a Cantor set--like structure typical of fractal sets. This structure is maintained if the scale is reduced, see Fig.  \ref{fig:1a}b, which is a feature of scale--invariance and self--similarity. The same can be observed for the dynamic landscape (Fig.  \ref{fig:1a}c,d). Here, it can be seen that for small values of $\kappa$, the landscape has no fractal features. This is due to the fact that for small target times $\kappa$, targeting by small perturbations of the parameters cannot be achieved generically.  For larger values of $\kappa$, approximately for $\kappa>20$, targeting is possible, and the landscape has fractal features. This is confirmed by  the box--counting dimension $D_{BC}$ of the landscapes, which is given in Fig. \ref{fig:2} for all four example systems.  The results show that for $\kappa$ below a certain threshold, the dimension is equal to $2$, which corresponds with the target time not large enough to achieved targeting. As the target time $\kappa$ gets larger, the dimension also increases and is no longer an integer.  The value of $D_{BC}$ converges to a constant for $\kappa$ getting large enough (approximately $\kappa >30$), which is almost the same  for all four tested landscapes (about $D_{BC} \approx 2.6$). Hence, it can be concluded that there is no direct relation between the fractal dimension of the chaotic attractors of the studied maps (see Tab. \ref{tab:1}) and the fractal dimension of the fitness landscapes that results from employing the maps to solve the targeting problem. Further note that there is a difference in the value of $D_{BC}$ depending on the initial edge length $\delta_0$, where smaller values of  $\delta_0$ produce slightly higher values of $D_{BC}$. However, this effect diminishes and finally comes to an end by decreasing $\delta_0$ further.  This is consistent as  the calculation of $D_{BC}$ is done for different values of $\delta_i$, if different values of $\delta_0$ are taken, see also Appendix A.

The next set of experiments are presented in Fig. \ref{fig:3} and give the landscape measures correlation length $\lambda$ and information content $h_{IC}$ over target time $\kappa$. We vary the scale on which the landscape measures are calculated by $\eta=0.01$ (solid lines), $\eta=0.0001$ (dashed lines), and $\eta=0.000001$ (dotted line), and again look at the effect of changing the initial edge length $\delta_0$.  Large values of the correlation length $\lambda$ indicate strong correlations between different regions of the landscape and hence weak ruggedness, while small values of $\lambda$ identify weak correlations and hence strong ruggedness. For the information content there is a direct relationship between the values of $h_{IC}$ and ruggedness.   
The results for both $\lambda$ and $h_{IC}$ allow to differentiate between small values of $\kappa$ and large values of $\kappa$ in terms of ruggedness measures, where for small $\kappa$ ruggedness is weaker and for large $\kappa$ ruggedness is stronger. This confirms that fractality is directly linked to ruggedness as the dimension of the landscapes is fractal for larger values of $\kappa$, as reported in Fig.    \ref{fig:2}. It can be further noted that this property holds if the scale of the landscape is reduced. In other words, fractal landscapes do not get smoother as the scale gets smaller. However, we see the effect that the transition between weak indication of ruggedness to strong indication is shifted to larger values of the target time $\kappa$ for $\eta$ getting smaller. The more we reduce the scale, the longer it takes until ruggedness takes shape on this scale. This effect is not visible in calculating  fractal dimensions as computing $D_{BC}$ is based on averaging over varying scales (see also appendix A), while the landscape measures are calculated for each scale. The transitions between weak and strong ruggedness vary over the different example systems with the coupled logistic map (\ref{eq:couplog}) producing the largest shift, see Fig. \ref{fig:3}h where the graphs for $\eta=0.000001$ are not depicted as they have values larger than zero for target times $\kappa > 50$. Furthermore, the experimental results suggest that the fractal dimension of the chaotic attractors of the studied maps (see Tab. \ref{tab:1} for the Kaplan--Yorke dimension $D_{KY}$ and the box--counting dimension $D_{BC}$) may be related to the transitions between weak and strong ruggedness over varying target time $\kappa$. The coupled logistic map (\ref{eq:couplog}) not only produces the largest shift, but also possesses the largest fractal dimension of all studied maps. On the other hand, the chaotic attractors of the H\'enon map (\ref{eq:henon}) and the Holmes map (\ref{eq:holmes}) have the smallest fractal dimension and also the smallest shift. The exponential map (\ref{eq:genexp2}) is in--between. It would be interesting to address in further studies if this relation holds for other maps as well, particularity such with higher dimension.   Another observation is that for small $\kappa$ the correlation length $\lambda$ has  largest values for initial edge length $\delta_0=\eta/256$ regardless of the value of $\eta$ and hence of the scale of the landscape. This means $\lambda$ indicates more smoothness for smaller values of $\delta_0$.  This is explainable by the fact that for small $\kappa$ no targeting can be achieved and small variation in $p$ does result in almost the same value of the distance function $j(p,\kappa)$. As scale reduces, the smoothening effect gets even stronger.  

The results show that the ruggedness measures correlation length $\lambda$ and information content $h_{IC}$ are scale--invariant and ruggedness is self--similar in fractal landscapes. By zooming in on the landscape, we do not reach a scale in which the landscape is smoother, but the same level of ruggedness persists. This property cannot be observed for non--fractal landscapes as to be seen for small target times $\kappa$. It can be expected that other ruggedness measures show similar characteristics. Furthermore, it can be seen  that box--counting dimension and both correlation length and information content are well related over target time $\kappa$. An exception from this rule  are the results for the coupled logistic map (\ref{eq:couplog}).  It might be assumed that the reason for this exception is the dynamic behavior of the map (\ref{eq:couplog}), as it is hyperchaotic for the parameter values considered and forms an chaotic attractor of very large fractal dimension.  Further work should be done to clarify this assumption.

\section{Concluding remarks} \label{sec:con}
Chaos can be used to direct trajectories to targets on a chaotic attractor. The targeting problem has been formulated as a dynamic fitness landscape and the resulting landscape has been analyzed with respect to ruggedness and fractality. The main findings of this analysis are that fractality can be seen as an upper limit of ruggedness and that landscape measures such as correlation length and information content evaluating ruggedness show self--similarity and other fractal properties by varying the scales upon which they are calculated. For fractal landscapes ruggedness does not come to an end if the scale of the configuration space variables gets small. 
Fitness landscapes have been defined to be fractal if the  Hausdorff--Besicovitch dimension of the fitness distribution over the configuration space is larger than the topological dimension of the configuration space. The results show that an equivalent definition of fractality is that landscape measures are invariant with respect to their scales.  

The results reported in this paper are for self--similar landscapes of parameter perturbations to achieve targeting. As targeting can also be realized by perturbations of the initial states, it might be interesting to analyze the
initial state landscapes of targeting. This would straightforwardly extend the results to higher dimensional targeting landscapes, particularly if we consider maps of higher dimension.

\section*{Appendix A: Fractal dimensions and box--counting}
Sets that lie in an $m$--dimensional Euclidean space may have a fractal dimension, which means that the dimension is not an integer. The main idea of assigning a fractal dimension to the set by box--counting is to cover the $m$--dimensional space by boxes, which have  dimension $m$ and edge length $\delta$. Hence, if $m=1$, the boxes are intervals of length $\delta$, while for $m=2$, the boxes are squares of area $\delta^2$. Choose an edge length $\delta$ and count the smallest number of boxes $N(\delta)$ that is needed to completely cover the set.  Repeat the counting for smaller and smaller $\delta$. The box--counting dimension is \begin{equation} D_{BC} = \underset{\delta \rightarrow
0} {\lim} \: \frac{\ln N(\delta)}{\ln (1/\delta)}.\end{equation}
We apply box--counting to calculating fractal dimensions of dynamic landscapes (\ref{eq:opttargetdyn}) using known computational techniques~\cite{block90,molt93}. A consequence of calculating $D_{BC}$ is that the configuration space (\ref{eq:limit}) is partitioned by boxes of edge length $\delta$. As the $p \in \mathcal{P}$ are bounded by $\eta$ and centered about the nominal value $\bar{p}$, we obtain a finite and pre--determined number of boxes that cover the landscape. The calculation of $D_{BC}$ starts with partitioning the  configuration space bounded by $\eta$ with boxes of edge length $\delta_0$ and  count the number $N(\delta_0)$. Halve $\delta_0$ to get $\delta_1$  and recount $N(\delta_1)$. By repeating the process $j$ times, we obtain $j$ pairs $(\delta_i, N(\delta_i))$.
The $D_{BC}$ is finally computed by the least squares linear fit of $\ln N(\delta_i)$ versus $\ln (1/\delta_i)$.

\section*{Appendix B: Numerical ruggedness measures: Correlation length and information content}
The landscape measures correlation length $\lambda$ and information content $h_{IC}$ are calculated by processing the 
sequence (\ref{eq:seq}).  For the  correlation length $\lambda$, the autocorrelation of (\ref{eq:seq}) with time lag $t_L$  yields the landscape's random walk correlation function
\begin{equation} r(t_L)=\frac{\underset{i=0}{\overset{T-1-t_L}{\sum}} \left(
f(p_i)-\mu\right) \left(f(p_{i+t_L})-\mu
\right) }{\sigma^2},\end{equation}
with
$\mu=\frac{1}{T}\underset{i=0}{\overset{T-1}{\sum}}
f(p_i)$, $\sigma^2 =\frac{1}{T} \underset{i=0}{\overset{T-1}{\sum}}
\left(f(p_i)-\mu\right)^2$ and $T \gg t_L>0$.  The
function $r(t_L)$ measures the correlation between different
regions of the fitness landscape and defines a measure of how smooth or rugged the landscape is. 
 The correlation length \begin{equation} \lambda=-1/\log(|r(1)|) \label{eq:corrle}\end{equation} derives from the autocorrelation  $r(1)$ of time lag $t_L=1$,~\cite{stad96,richengel14}.

For calculating the information content $h_{IC}$,  we take  the sequence (\ref{eq:seq}) and code
differences in fitness  between two consecutive walking steps by
symbols $s_{i} \in \mathbb{S}$, $i=0,1,2,\ldots,T-1$,
taken from the set $\mathbb{S}= \{-1,0,1\}$. We obtain  these symbols 
 by
\begin{equation}s_i(\epsilon)=\left\{ \begin{array}{rcccc} -1,
& \quad \text{if} \quad& f(p_{i+1})-f(p_i) & < & \epsilon \\ 0, & \quad \text{if} \quad &  |f(p_{i+1})-f(p_i)|&\leq & \epsilon\\
1, & \quad \text{if}  \quad& f(p_{i+1})-f(p_i)&> & \epsilon
\end{array} \right.
\end{equation}
for a fixed $\epsilon \in [0,L]$, where $L$ is the maximum difference
between two fitness values.   Concatenating the symbols $s_i$ gives the string
\begin{equation} S_{IC}(\epsilon)=s_0s_1\ldots s_{T-1}. \label{eq:entstring} \end{equation} The sensitivity level $\epsilon$ defines the accuracy with which the string
$S_{IC}(\epsilon)$ accounts for differences in the fitness values. For example, 
if $\epsilon=0$, the string $S_{IC}(\epsilon)$ contains the symbol zero only for the
random walk reaching a strictly flat area. Hence, $\epsilon=0$
discriminates very sensitively between increasing and decreasing
fitness values. If, on the other hand, $\epsilon=L$, the string only
contains the symbol zero, which makes evaluating the structure of
the landscape meaningless. A fixed value of $\epsilon$ with
$0<\epsilon<L$ defines a level of detail with respect to the information gained about the
landscape structure

For defining the information content of the landscape, the
distribution of subblocks of length two, $s_{i}
s_{i+1}$, $i=0,1,\ldots T-2$, within the string
(\ref{eq:entstring}) is analyzed. These subblocks stand for local patterns in
the landscape. The probabilities of the occurrence of the
pattern $ab$ with $a,b \in
\mathbb{S}$ and $a \neq b$ are denoted by $p_{ab}$. For numerical calculation,  these
probabilities are approximated by the relative frequency of the patterns within the
string (\ref{eq:entstring}). As the set $\mathbb{S}$ consists of three
elements, we find $6$ different kinds of subblock $s_{i}
s_{i+1}=ab$ with $a \neq b$
within the string $S_{IC}(\epsilon)$. From their probabilities
and a given sensitivity level $\epsilon$  the entropic
measure
\begin{equation} h_{IC}(\epsilon)=- {\underset{a, b \in
\mathbb{S} \atop a \neq b}{\sum}} p_{ab}  \log_6  p_{ab},
\label{eq:infcont}
\end{equation}
is calculated, which is called  information content of the fitness landscape,~\cite{mun15,vassi00}.
Note that by taking the logarithm in Eq. (\ref{eq:infcont}) with
the base $6$, the information content is scaled to the interval
$[0,1]$.


\label{lastpage}


\begin{thebibliography}{12}
\markboth{Hendrik Richter}{The International Journal of Parallel, Emergent and Distributed Systems}

\bibitem{alv00} N.~A.~Alves,  U.~H.~E.~Hansmann. Glass transition temperature and fractal dimension of protein free energy landscapes.  ‎Int. J. Mod. Phys. C 11,  301--308, 2000.


\bibitem{ban05} N.~K.~Banavali, B. Roux,  Free energy landscape of A--DNA to B--DNA conversion in aqueous solution, J. Amer. Chem, Soc.  127, 6866--6876, 2005. 

\bibitem{block90} A. Block,  W. von Bloh,  H. J. Schellnhuber. Efficient box--counting determination of generalized fractal dimensions. Phys. Rev. A42,  1869--1874, 1990.


\bibitem{boll05} E. M. Bollt,   The path towards a longer life: On invariant sets and the escape time landscape. Int. J.  Bifurcation and Chaos 15, 1615--1624, 2005.

\bibitem{brede11}  M. Brede, The evolution of cooperation on correlated payoff landscapes. Artificial Life 17, 365--373, 2011.



\bibitem {corless96} R. M. Corless,  G. H.  Gonnet,   D. E. G.  Hare, D. J.
Jeffrey, D. E.  Knuth,  On the Lambert $\mathcal{W}$ function.
Advances Comput. Math. 5, 
 329--359, 1996.

\bibitem {ding96} M. Ding, W. Yang, Observation of intermingled
basins in coupled oscillators exhibiting synchronized chaos. Phys.
Rev. E54,  2489--2494, 1996.

 \bibitem{falc90} K.~J.~Falconer, Fractal Geometry: Mathematical Foundations and Applications. John Wiley \& Sons, Chichester, 1990.

\bibitem{gav04} S.  Gavrilets,    Fitness Landscapes and the Origin of Species.   Princeton University Press, Princeton, NJ, 2004.

\bibitem{gia02} L.~Giada,  M.~Marsili. Algorithms of maximum likelihood data clustering with applications. Physica A315,   650--664, 2002.

\bibitem{groe12} A. Gr{\"o}nlund, I. G. Yi, B. J.  Kim, Fractal profit landscape of the stock market. PloS one, 7(4), e33960, 2012.

\bibitem{hen76}  M. H\'enon, A two-dimensional mapping with a strange
attractor.  Commun. Math. Phys. 50, 69--77, 1976.

\bibitem{holm79} P.~J.~Holmes, A nonlinear oscillator with a strange attractor.
Philos. Trans. R. Soc. London  A 292, 419--448, 1979.

\bibitem{hosh98} T. Hoshino,  D. Mitsumoto,  T.  Nagano, Fractal fitness landscape and loss of robustness in evolutionary
robot navigation. Autonomous Robots 5, 199--213, 1998.

\bibitem{ip02}   S. Iplikci, Y. Denizhan, Targeting in dissipative chaotic systems: A survey.
Chaos 12, 995--1005, 2002. 

\bibitem{kap79} J.~L. Kaplan,  J.~A. Yorke: Chaotic behavior
in multidimensional difference equations. In: H.~O. Peitgen, H.~O.
Walther (eds.), Functional Difference Equations and Approximations
of Fixed Points.  Springer--Verlag, Berlin,  204--227, 1979.

\bibitem{kauf87} S.~A.~Kauffman, S. Levin, Towards a general theory of adaptive walks on rugged landscapes.  J. Theor. Biology 128, 11--45, 1987.

\bibitem{leit13} J.~C.~Leit\~{a}o,  J.~M.~Viana Parente Lopes,  E.~G.~Altmann, Monte Carlo sampling in fractal landscapes.  Phys.  Rev. Lett. 110, 220601, 2013.




\bibitem{lloyd95} A. L. Lloyd,   The coupled logistic map: a simple model for the effects of spatial heterogeneity on population dynamics. J. Theor. Biology 173, 217--230, 1995.

\bibitem{mac06} C.  MacNish,  Benchmarking evolutionary and hybrid algorithms using randomized self--similar landscapes. In:  T. D. Wang et al. (eds.), Proc.  Asia--Pacific Conference on Simulated Evolution and Learning.  Springer--Verlag,  Berlin Heidelberg,   361--368, 2006.

\bibitem{man83} B. B. Mandelbrot,  The Fractal Geometry of Nature. Freeman, New York, 1983.

\bibitem{molt93} T.~C.~A.~Molteno,  Fast $\mathcal O(N)$ box--counting algorithm for estimating dimensions. Phys. Rev. E48,  R3263--3266, 1993.

\bibitem{mun15}  M. A. Mu\~noz, M. Kirley,  S. K.  Halgamuge,  Exploratory landscape analysis of continuous space optimization problems using information content.  IEEE Trans. Evolut. Comp.  19, 74--87, 2015.


\bibitem{on97} J.~N.~Onuchic, Z. Luthey Schulten, P.~G.~Wolynes, Theory of protein folding: The energy landscape perspective,  Ann. Rev. Phys. Chem. 48, 545--600, 1997.

\bibitem{palmer91} R. Palmer: Optimization on rugged landscapes. In:  A. S. Perelson, S. A. Kauffman (eds.),
 Molecular Evolution on
Rugged Landscapes: Proteins, RNA, and the Immune System, Addison Wesley, Redwood City, CA, 
3--25, 1991.


\bibitem {pas95} M. Paskota, A. I. Mees, K. L.  Teo, Geometry of targeting of chaotic systems.
Int. J. Bifurcation and Chaos 5, 1167-1173, 1995.


 \bibitem{palao13} J. P. Palao,  D. M. Reich, C. P.  Koch,  Steering the optimization pathway in the control landscape using constraints. Phys. Rev. A88, 053409, 2013.

\bibitem{rabitz14} H. Rabitz, R. B. Wu, T. S. Ho, K. M. Tibbetts, X.  Feng,  Fundamental principles of control landscapes with applications to quantum mechanics, chemistry and evolution.   In: H. Richter, A. P.  Engelbrecht (eds.), Recent Advances in the Theory and Application of Fitness Landscapes, Springer--Verlag, Berlin Heidelberg New York,  33--70, 2014.

\bibitem {rich08} H.~Richter,
Coupled map lattices as spatio--temporal fitness functions:
Landscape measures and evolutionary optimization.  Physica
 D237,  167--186, 2008.


\bibitem {rich08a} H.~Richter, Can a polynomial interpolation improve on the Kaplan--Yorke dimension? Phys. Lett. A372,  4689--4693, 2008.

\bibitem{richt08}  H.~Richter, On a family of maps with multiple chaotic attractors. Chaos, Solitons \& Fractals 36, 559--571, 2008.

\bibitem{rich12} H. Richter, Analyzing dynamic fitness landscapes of the targeting problem of chaotic systems. In: C. Di Chio et al. (eds.),  Applications of Evolutionary Computation - EvoApplications 2012, Springer-Verlag, Berlin,  83--92, 2012.


\bibitem{rich14a} H.~Richter,  Fitness landscapes that depend on time. In: H. Richter, A. P.  Engelbrecht (eds.), Recent Advances in the Theory and Application of Fitness Landscapes, Springer--Verlag, Berlin Heidelberg New York,  265--299, 2014.





\bibitem{richengel14} H. Richter, A. P. Engelbrecht,  Recent Advances in the Theory and Application of Fitness Landscapes.  Springer-Verlag, Berlin, 2014. 


\bibitem{sork91}  G.  B.~Sorkin,   Efficient simulated annealing on fractal energy landscapes. Algorithmica 6, 367--418, 1991.



    \bibitem {stad96} P.~F.~Stadler,  Landscapes and their correlation
functions. J. Math. Chem. 20,  1--45, 1996.

\bibitem {vall00} S.~R. Valluri, D.~J. Jeffrey, R.~M.  Corless, Some
applications of the Lambert $\mathcal{W}$ function to physics. Can. J.
Phys. 78,  823--831, 2000.

\bibitem{vassi00}  V.~K. Vassilev, T.~C. Fogarty,  J.~F.  Miller, 
    Information characteristics and the structure of landscapes. Evolut.
    Comput. 8,  31--60, 2000.

\bibitem{vonbre97} H.~F. von Bremen, F.~E. Udwadia, W. Proskurowski, An efficient QR based method for the computation of Lyapunov exponents.  Physica D101, 1--16, 1997.

\bibitem{wal04}  D. J.  Wales, Energy Landscapes: Applications to Clusters, Biomolecules and Glasses.  Cambridge University Press, Cambridge, UK, 2004.



\bibitem {wein90}  E.~D.~Weinberger,
Correlated and uncorrelated fitness landscapes and how to tell the
difference.  Biol. Cybern.  63,  325--336, 1990.


 \bibitem{wein93} E.~D. Weinberger, P.~ F. Stadler, Why some fitness landscapes are fractal. J.  Theor. Biology  163, 255--275, 1993.



\bibitem{zelink14} I. Zelinka,  O. Zmeskal, P. Saloun. Fractal analysis of fitness landscapes. In: H. Richter, A. P.  Engelbrecht (eds.), Recent Advances in the Theory and Application of Fitness Landscapes, Springer--Verlag, Berlin Heidelberg New York, 427--456, 2014.






\end{thebibliography}
\end{document}